\documentclass[acmsmall,screen]{acmart}
\usepackage[T1]{fontenc}
\usepackage[utf8]{inputenc}
\usepackage{fontawesome5}

\usepackage{mathpartir}
\usepackage{graphicx}
\usepackage{hyperref}
\usepackage{multicol}
\usepackage{stmaryrd} %used by iris
\usepackage{iris}
\usepackage{xspace}
\usepackage{xcolor}
\usepackage{amsfonts}
\usepackage{adjustbox}
\usepackage{soul}

\newenvironment{DIFnomarkup}{}{}

\newcommand{\captionlabel}[2]{%
  \vspace{-1em}%
  \caption{#1}%
  \Description{#1}%
  \label{#2}
  \vspace{-1em}%
}

\newcommand\lang{MusketLang\xspace}
\newcommand{\musketeer}{Musketeer\xspace}
\newcommand{\chainedlog}{ChainedLog\xspace}
\newcommand{\angelic}{Angelic\xspace}
\newcommand{\heaplang}{HeapLang\xspace}
\newcommand{\typesystem}{MiniDet\xspace}
\newcommand{\minidet}{\typesystem}

\newcommand{\Values}{\mathcal{V}}
\newcommand{\val}{v}
\newcommand{\vals}{\vec{\val}}
\newcommand{\vunit}{()}
\newcommand{\loc}{\ell}
\newcommand{\Loc}{\mathcal{L}}
\newcommand{\vint}{i}
\newcommand{\vbool}{b}
\newcommand{\vtrue}{\mathsf{true}}
\newcommand{\vfalse}{\mathsf{false}}
\newcommand{\vfun}[3]{\hat\mu#1\,#2.\,#3}
\newcommand{\vprod}[2]{(#1,#2)}

\newcommand{\primitive}{\bowtie}
\newcommand{\symbeq}{\mathord{==}}
\newcommand{\symbor}{\lor}
\newcommand{\symband}{\land}

\newcommand{\expr}{e}
\newcommand{\eassert}[1]{\mathsf{assert}\,#1}
\newcommand{\elet}[3]{\mathsf{let}\,#1\,=\,#2\,\mathsf{in}\,#3}
\newcommand{\efun}[3]{\mu#1\,#2.\,#3}
\newcommand{\efunnonrec}[2]{\lambda#1.\,#2}
\newcommand{\eref}[1]{\mathsf{ref}\,#1}
\newcommand{\eget}[1]{\mathsf{get}\,#1}
\newcommand{\eset}[2]{\mathsf{set}\,#1\,#2}
\newcommand{\ecall}[2]{#1\,#2}
\newcommand{\epar}[2]{\mathsf{par}\,#1\,#2}
\newcommand{\erunpar}[2]{#1\,||\,#2}
\newcommand{\eprod}[2]{\mathsf{prod}\,#1\,#2}
\newcommand{\ealloc}[1]{\mathsf{alloc}\;#1}
\newcommand{\eload}[2]{#1[#2]}
\newcommand{\estore}[3]{#1[#2]\!\leftarrow\! #3 }
\newcommand{\elength}[1]{\mathsf{length}\,#1}
\newcommand{\eif}[3]{\mathsf{if}\,#1\,\mathsf{then}\,#2\,\mathsf{else}\,#3}
\newcommand{\ecas}[4]{\mathsf{CAS}\,#1\,#2\,#3\,#4}
\newcommand{\eeq}[2]{#1\,\symbeq\,#2}
\newcommand{\blockrepeat}[2]{#2^{#1}}
\newcommand{\eapp}[2]{#1\,#2}
\newcommand{\ecallprim}[2]{#1\primitive#2}
\newcommand{\eproj}[2]{\mathsf{proj}_{#1}\,#2}

\newcommand{\eletprefix}[2]{\textsf{let}\,#1\,=\,#2\,\textsf{in}}

\newcommand{\store}{\sigma}
\renewcommand{\subst}[3]{[#2/#1]#3}
\renewcommand{\dom}[1]{\mathsf{dom}(#1)}
\newcommand{\blockupd}[3]{[#2:=#3]#1}
\newcommand{\wal}{w}
\newcommand{\wals}{\vec\wal}
\newcommand{\length}[1]{|#1|}

\newcommand{\ectx}{K}
\newcommand{\efillctx}[2]{#1\langle#2\rangle}
\newcommand{\khole}{\square}

\newcommand{\config}[2]{#1\,\backslash\,#2}

\newcommand{\petit}[1]{\text{\scriptsize\rm\sf#1}}
\newcommand{\steprightarrowrule}[1]
{\mathrel{\xrightarrow{\adjustbox{margin=0ex 0ex 0ex
        1.2ex}{\raisebox{-0.2ex}[0ex][0ex]{#1}}}}}
\newcommand{\purestep}[2]{#1 \;\steprightarrowrule{\petit{pure}}\; #2}
\newcommand{\headstep}[4]{\config{#1}{#2} \,\steprightarrowrule{\petit{head}}\, \config{#3}{#4}}
\newcommand{\step}[4]{\config{#1}{#2} \,\longrightarrow\, \config{#3}{#4}}

\newcommand{\steprtc}[4]{\config{#1}{#2} \,\longrightarrow^{\ast}\, \config{#3}{#4}}

\newcommand{\triple}[3]{\{#1\}\;#2\;\{#3\}}

\newcommand{\atriple}[4]{\triple{#1}{#2}{\lambda#3\,\_.\;#4}}
\newcommand{\atriplemore}[5]{\triple{#1}{#2}{\lambda#3\,#4.\;#5}}

\newcommand{\mystackrel}[2]{#2_{#1}}
\newcommand{\prepointsto}{\mapsto}
\newcommand{\pointsto}{\mathrel{\prepointsto}}
\newcommand{\fpointsto}[1]{\mathrel{\mystackrel{#1}{\prepointsto}}}

\newcommand{\pointstox}[1]{\prepointsto^{#1}}

\renewcommand{\star}{\ast}
\newcommand{\wider}[1]{\,#1\,}
\newcommand{\morespacingaroundstar}{%
\let\oldstar\star
\renewcommand{\star}{\wider\oldstar}%
}
\newcommand{\morespacingaroundwedge}{%
\let\oldwedge\wedge
\renewcommand{\wedge}{\wider\oldwedge}%
}
\newcommand{\pure}[1]{\ulcorner #1 \urcorner}

\newcommand{\iTrue}{\top}
\newcommand{\pre}{\varphi}
\newcommand{\post}{\psi}

\newcommand{\vpre}{P}
\newcommand{\vpost}{Q}

\newcommand{\qp}{q}

\newcommand{\ofs}{i}

\newcommand{\ctriplebase}[6]{\{#1\}\;#2\;\{#3 \mid #4\}\;#5\;\{#6\}}
\newcommand{\ctriple}[8]{\ctriplebase{#1}{\efillctx{#2}{#3}}{#4}{#5}{\efillctx{#6}{#7}}{#8}}
\newcommand{\gol}{\pmb{l}}
\newcommand{\gor}{\pmb{r}}
\newcommand{\ectxs}{\vec{\ectx}}
\renewcommand{\wp}[3]{\textsf{wp}_{#1}\;#2\;\{#3\}}

\newcommand{\iProp}{\ensuremath{\mathit{iProp}}}
\newcommand{\vProp}{\ensuremath{\mathit{vProp}}}

\newcommand{\bigast}[2]{{\scaleobj{2}{\ast}}_{#1}\,#2}

\newcommand{\tid}{\pi}

\newcommand{\yielded}[2]{\textsf{yielded}\,#1\,#2}

\newcommand{\goal}{\mathfrak{goal}}
\newcommand{\runname}{\textsf{run}\xspace}
\newcommand{\run}[2]{\runname\,#1\,\{#2\}}
\newcommand{\runex}[3]{\run{#1}{\lambda#2.\,#3}}

\newcommand{\fpointstox}[2]{\prepointsto^{#1}_{#2}}
\newcommand{\leftalloc}[1]{\textsf{leftalloc}\,#1}

\newcommand{\judg}[4]{#1 \,\vdash\,#2\,:#3\,\dashv\,#4}

\newcommand{\msingleton}[2]{\{#1:=#2\}}
\newcommand{\minsert}[3]{[#1:=#2]#3}
\newcommand{\mdelete}[2]{\textsf{del}\,#1\,#2}

\newcommand{\tunit}{\textsf{unit}}
\newcommand{\tbool}{\textsf{bool}}
\newcommand{\tint}{\textsf{int}}
\newcommand{\tarrow}[2]{#1 \rightarrow #2}
\newcommand{\tref}[1]{\textsf{ref}\,#1}

\newcommand{\tpwrite}[1]{\textsf{pwrite}\,#1}
\newcommand{\tpread}[1]{\textsf{pread}\,#1}
\newcommand{\tprod}[2]{(#1 \times #2)}

\newcommand{\pallocname}{\textsf{palloc}}
\newcommand{\palloc}[1]{\ecall{\pallocname}{#1}}

\newcommand{\pwritename}{\textsf{pwrite}}
\newcommand{\pwrite}[2]{\ecall\pwritename\,#1\,#2}

\newcommand{\preadname}{\textsf{pread}}
\newcommand{\pread}[1]{\ecall{\preadname}{#1}}

\newcommand{\sltriple}[3]{\triple{#1}{#2}{#3}_{SL-}}

\newcommand{\motto}[2]{if one execution of $#1$ is safe and terminates, then every execution of $#2$ is safe\xspace}

\newcommand{\indirection}{\textsf{indirection}}

\newcounter{remark}[section]

\usepackage{xspace}
\usepackage{expl3}
\ExplSyntaxOn
\newcommand\latinabbrev[1]{
  \peek_meaning:NTF . {% Same as \@ifnextchar
    \emph{#1}\@}%
  { \peek_catcode:NTF a {% Check whether next char has same catcode as \'a, i.e., is a letter
      \emph{#1}.\@\xspace}%
    {\emph{#1}.\@\xspace}}}
\ExplSyntaxOff

\def\eg{\latinabbrev{e.g}}

\def\ie{\latinabbrev{i.e}}

\newcommand{\allocfillname}{\textsf{alloc\_fill}\xspace}
\newcommand{\allocfill}[2]{\allocfillname\,#1\,#2}

\newcommand{\hinitname}{\textsf{init}\xspace}
\newcommand{\hinit}[2]{\hinitname\,#1\,#2}

\newcommand{\haddname}{\textsf{add}\xspace}
\newcommand{\hadd}[2]{\haddname\,#1\,#2}
\newcommand{\helemsname}{\textsf{elems}\xspace}
\newcommand{\helems}[1]{\helemsname\,#1}
\newcommand{\filtercompactname}{\textsf{filter\_compact}\xspace}
\newcommand{\fillname}{\textsf{fill}\xspace}

\newcommand{\SL}{separation logic\xspace}

\renewcommand{\TirNameStyle}[1]{\hypertarget{#1}{\textsc{#1}}}

\newcommand{\exname}{\textsf{dumas}\xspace}
\newcommand{\aaddname}{\textsf{atomic\_add}\xspace}
\newcommand{\aadd}[2]{\aaddname\,#1\,#2}

\newcommand{\deduped}[2]{\textsf{deduped}\,#1\,#2}

\newcommand{\sintarray}[1]{\textsf{sintarray}\,#1}

\usepackage[nameinlink,noabbrev]{cleveref}
\crefname{section}{\S\!}{\S}
\Crefname{section}{Section}{Sections}
\crefname{figure}{Figure}{Figures}
\crefname{theorem}{Theorem}{Theorems}

\newtheorem{theorem}{Theorem}[section]

\newcommand{\citeappendix}[1]{the Appendix~\citep{extendedversion}}

\newif\ifappendix
\appendixtrue % UNCOMMENT TO GET THE APPENDIX
\ifappendix
\renewcommand{\citeappendix}[1]{\Cref{#1}}
\fi

\renewcommand{\TirNameStyle}[1]{\hypertarget{#1}{\textsc{#1}}}
\newcommand{\RULE}[1]{\hyperlink{#1}{\textsc{#1}}\xspace}
\setcitestyle{nosort}

\begin{document}

\title{All for One and One for All: Program Logics for Exploiting Internal Determinism in Parallel Programs}

\ifappendix
\subtitle{Extended Version}
\fi

\author{Alexandre Moine}
\orcid{0000-0002-2169-1977}
\email{alexandre.moine@nyu.edu}
\affiliation{%
  \institution{New York University}
  \city{New York}
  \country{USA}
}

\author{Sam Westrick}
\orcid{0000-0003-2848-9808}
\email{shw8119@nyu.edu}
\affiliation{%
  \institution{New York University}
  \city{New York}
  \country{USA}
}

\author{Joseph Tassarotti}
\orcid{0000-0001-5692-3347}
\email{jt4767@nyu.edu}
\affiliation{%
  \institution{New York University}
  \city{New York}
  \country{USA}
}

%%% The following is specific to  and the paper
%%% 'All for One and One for All: Program Logics for Exploiting Internal Determinism in Parallel Programs'
%%% by Alexandre Moine, Sam Westrick, and Joseph Tassarotti.
%%%
\setcopyright{cc}
\setcctype{by}
\acmDOI{10.1145/3776668}
\acmYear{2026}
\acmJournal{PACMPL}
\acmVolume{10}
\acmNumber{POPL}
\acmArticle{26}
\acmMonth{1}
\received{2025-07-10}
\received[accepted]{2025-11-06}

\begin{abstract}
Nondeterminism makes parallel programs challenging to write and reason
about. To avoid these challenges, researchers have developed
techniques for internally deterministic parallel programming, in which
the steps of a parallel computation proceed in a deterministic way.
Internal determinism is useful because it lets a programmer reason
about a program as if it executed in a sequential order. However, no
verification framework exists to exploit this property and simplify
formal reasoning about internally deterministic programs.

To capture the essence of why internally deterministic programs should
be easier to reason about, this paper defines a property called
schedule-independent safety. A program satisfies schedule-independent
safety, if, to show that the program is safe across all orderings, it
suffices to show that one terminating execution of the program is
safe. We then present a separation logic called Musketeer for proving
that a program satisfies schedule-independent safety. Once a parallel
program has been shown to satisfy schedule-independent safety, we can
verify it with a new logic called Angelic, which allows one to
dynamically select and verify just one sequential ordering of the
program.

Using Musketeer, we prove the soundness of MiniDet, an affine type
system for enforcing internal determinism. MiniDet supports several
core algorithmic primitives for internally deterministic programming
that have been identified in the research literature, including a
deterministic version of a concurrent hash set. Because any
syntactically well-typed MiniDet program satisfies
schedule-independent safety, we can apply Angelic to verify such
programs.

All results in this paper have been verified in Rocq using the Iris
separation logic framework.

\end{abstract}


% DIFnomarkup is an empty env (defined above) that removes markup from latexdiff
\begin{DIFnomarkup}
\end{DIFnomarkup}

\maketitle

\section{Introduction}
\label{sec:intro}
One of the most challenging aspects of concurrent and parallel programming is dealing with nondeterminism.
Nondeterminism complicates almost every aspect of trying to make programs correct.
Bugs often arise because programmers struggle to reason about the set of all possible nondeterministic outcomes and interleavings.
Finding those bugs becomes more difficult, as testing can only cover a subset of possible outcomes.
Even when bugs are found, nondeterminism makes them harder to reproduce and debug.
%In addition, nondeterminism makes static analysis and verification more complicated.
These challenges also extend to formal methods for such programs, where nondeterminism makes various analyses and verification techniques more complex.

For these reasons, there has long been interest in methods for \emph{deterministic} parallel programming.
A range of algorithmic techniques~\citep{bfgs12-pbbs}, language designs~\citep{DBLP:journals/jpdc/BlellochHSZC94, DBLP:conf/pldi/KuperTTN14}, type systems~\citep{DBLP:conf/oopsla/BocchinoADAHKOSSV09}, specialized operating systems and runtimes~\citep{DBLP:conf/osdi/AviramWHF10}, and various other approaches have been developed for making parallel programs deterministic.
Researchers in this area have long noted that determinism is not simply a binary property, and in fact there is a spectrum of \emph{degrees} of determinism.
On one end of the spectrum is \emph{external determinism}, which simply says that the input/output behavior of a program is deterministic.
However, in an externally deterministic program, even if the final output is deterministic, the manner in which the computation takes place may be highly nondeterministic and vary across runs.
As a result, external determinism does not eliminate all of the programming challenges associated with nondeterminism.
For example, a programmer who attaches a debugger to an externally deterministic program may still see different internal behaviors across different runs, complicating efforts to understand the program's behavior.

A stronger property, called \emph{internal determinism}, requires in addition that the structure and internal steps of a computation are deterministic.
More formally, in an internally deterministic program, for a given input,
% to the program,
every execution will generate the same \emph{computation graph}, a
% kind of
trace that captures the dependencies of operations and their results.
With this strong form of determinism, we can reason about the program's behavior by considering any \emph{one} sequential traversal of operations in the computation graph.
This is useful, because as \citet{bfgs12-pbbs} put it:
\begin{quote}
In addition to returning deterministic results, internal determinism has many advantages including ease of reasoning about the code, ease of verifying correctness, ease of debugging, ease of defining invariants, ease of defining
good coverage for testing, and ease of formally, informally and experimentally reasoning about performance.
\end{quote}
Although ensuring internal determinism might seem expensive, \citet{bfgs12-pbbs} have shown that by using a core set of algorithmic techniques and building blocks, it is possible to develop fast and scalable internally deterministic algorithms for a range of benchmark problems.
\newline{}

In this paper, we explore the meaning and benefits of internal determinism from the perspective of program verification.
If one of the advantages of internal determinism is that it simplifies reasoning about programs, then it should be possible to exploit this property in the form of new reasoning rules in a program logic.
% for internally deterministic programs.
To do so, we first define a property we call \emph{schedule-independent safety}, which holds for a parallel program $e$ if, to verify that every execution of $e$ is safe (\ie never triggers undefined behavior or a failing assert), it suffices to prove that at least \emph{one} interleaving of operations in $e$ is terminating and safe.
Internal determinism implies schedule-independent safety, and it is this property that makes reasoning about internally deterministic programs simpler.
Schedule-independent safety recalls the motto of Dumas' Three Musketeers, ``all for one and one for all'': the safety of all interleavings amounts to the safety of one of them.
Building on this observation, we develop \musketeer, a separation logic for proving that a program satisfies schedule-independent safety.
Although \musketeer is formulated as a unary program logic, schedule-independent safety is a $\forall\forall$ \emph{hyperproperty}~\citep{DBLP:journals/jcs/ClarksonS10}, since it relates safety of any chosen execution of a program $e$ to all other executions of $e$.
Thus, to prove the soundness of \musketeer, we encode \musketeer triples into a new relational logic called \chainedlog.
In contrast to most prior relational concurrent separation logics, which are restricted to $\forall\exists$ hyperproperties, \chainedlog supports $\forall\forall$ hyperproperties using a judgement we call a \emph{chained triple}.
%This relational logic is itself novel, because most prior relational separation logics for concurrent programs are restricted to proving $\forall \exists$ hyperproperties, but schedule-independent safety is a $\forall\forall$ hyperproperty.
%While most prior relational separation logics for concurrent programs are restricted to proving $\forall\exists$ hyperproperties,
% To encode $\forall\forall$ hyperproperties, we develop a new program logic called \chainedlog, and encode \musketeer using \chainedlog.

Intuitively, the high-level reasoning rules of \musketeer restrict
the user to verify only internally-deterministic programs.
However, while internally deterministic programs always satisfy schedule-independent safety,
the converse is false: a program may be nondeterministic because it observes
actions from concurrent tasks, but it may do so without jeopardizing safety.
In order to verify such programs~(\cref{sec:prio}, \cref{sec:hash}),
we use the fact that \musketeer is defined in terms of the
more flexible and more complex \chainedlog,
and drop down to this low-level logic to conduct the proof.

We next explore how to exploit schedule-independent safety to simplify verification of programs.
To that end, we present a logic called \angelic that allows one to \emph{angelically} select and verify one sequential ordering of operations in a parallel program.
\angelic is sound to apply to programs that satisfy schedule-independent safety because the safety and termination of the one ordering verified during the proof will imply safety for all other executions.
This is in contrast to standard concurrent separation logics, in which one must consider \emph{all} possible orderings during a proof.
%What makes this logic sound to apply to programs that are schedule-independent safe is that the safety and termination of the one ordering verified during the course of the proof will imply safety for all other executions.
% any ordering gives rise to the same computation graph, so that it suffices to verify just one.

Using these logics, we verify a number of examples from the literature on internal determinism and related properties.
First, we show how to use \musketeer to prove properties about language-based approaches for enforcing internal determinism.
In particular, because \musketeer is a higher-order impredicative logic, \musketeer can encode logical relations models for type systems that are designed to enforce internal determinism.
We start by applying this to a simple ownership-based affine type system we call \minidet.
% using the logical relation to prove that all well-typed programs in \minidet satisfy schedule-independent safety.
The resulting logical relations model for \minidet shows that every well-typed program satisfies schedule-independent safety.
Next we use \musketeer to prove specifications for \emph{priority writes} and \emph{deterministic concurrent hash sets}, two of the core primitives that \citet{bfgs12-pbbs} use in several of their examples of internally deterministic algorithms.
Using these specifications, we extend \minidet and its logical relations model with typing rules for priority writes and hash sets, showing that schedule-independent safety is preserved.

Finally, putting these pieces together, we turn to parallel array deduplication, one of the example benchmark problems considered by \citet{bfgs12-pbbs}.
We first show that an implementation of the algorithm they propose for this problem can be syntactically-typed in \minidet, thereby showing that it is schedule-independent safe.
Next, we use \angelic to verify a correctness property for this algorithm.
Although the algorithm is written using a parallel for-loop that does concurrent insertions into a hash set, by using \angelic, we can reason \emph{as if} the parallel loop was a standard, sequential loop, thereby simplifying verification.

\paragraph{Contributions}
The contributions of this paper are the following:
\begin{itemize}
\item We identify schedule-independent safety as a key property of deterministic parallel \mbox{programs}.
\item We present \musketeer, a \SL for proving that a program satisfies schedule-independent safety,
meant to be used as a tool for proving automatic approaches correct.
\item We present \angelic, a \SL for proving that \emph{one} interleaving safely terminates.
\item We use \musketeer to verify properties of \minidet, an affine type system guaranteeing schedule-independent safety.
\item We verify that priority writes and a deterministic concurrent hash set satisfy
schedule-independent safety using \musketeer, and then use this property to verify
a deduplication algorithm using \angelic.
\item We formally verify all the presented results~\citep{mechanization}, including the soundness of the logics and the examples, in the Rocq prover using the Iris framework~\citep{iris}.
\end{itemize}

\section{Key Ideas}
\label{sec:key_ideas}
In this section, we first give a simple motivating example~(\cref{sec:key:example}),
describe some of the core concepts behind how \musketeer guarantees schedule-independent safety~(\cref{sec:key:musketeer}),
and conclude by showing some of the rules of \angelic that allow
for reasoning sequentially about a parallel program~(\cref{sec:key:angelic}).

\subsection{A Motivating Example}
\label{sec:key:example}

Our example program is named \exname and appears below:
\begin{align*}
%&\exname \;\eqdef\; \lambda f.\\
\exname \;\eqdef\;\; \lambda n.\ &\eletprefix{r}{\eref{0}}\\
%&\quad\epar{(\lambda\_.\;\aadd{r}{(f\,\vtrue)})}{(\lambda\_.\;\aadd{r}{(f\,\vfalse)})};\\
&\epar{(\lambda\_.\;\aadd{r}{1802})}{(\lambda\_.\;\aadd{r}{42})};\\
&\eassert{(\eget{r} == n)}
\end{align*}
The \exname program takes an argument $n$.
It first allocates a reference $r$ initialized to~$0$,
and then calls in parallel two closures,
one that atomically adds $1802$ to $r$,
and the other that atomically adds $42$.
%
% Such an atomic update can be implemented with a compare-and-swap (CAS) loop or other primitive atomic instruction.
After the parallel phase, the function
asserts that the content of $r$ is equal to $n$.

Imagine we wish to prove that $(\exname\,1844)$ is safe---that is,
for every interleaving, the program will never get stuck,
and in particular the assertion will succeed.
Of course, many existing concurrent separation logics can easily prove this.
In such logics, one can use an \emph{invariant} assertion to reason about the shared access to $r$ by the two parallel threads.
This invariant would ensure that, no matter which order the threads perform their additions, after both have finished $r$ will contain $1844$.

We propose an alternate approach that simplifies reasoning by exploiting the internal determinism in programs like $\exname$.
In our approach, we first prove in a light-weight way that, for any given value of $n$, the order of the parallel additions in $(\exname\,n)$ does not affect the outcome of the assert.
Then, to prove safety of $(\exname\, n)$ for the specific value of $n = 1844$, we can just pick \emph{one} possible ordering and verify safety of that ordering.

\subsection{Verifying Schedule-Independent Safety with Musketeer}
\label{sec:key:musketeer}

Our first contribution is \musketeer, a logic for proving that a program satisfies schedule-independent safety, \ie that safety of any one complete execution implies safety of all possible executions.
% In order to verify that a program satisfies schedule-independent safety,
% we propose \musketeer~(\cref{sec:logic}).
Although \musketeer is itself a program logic, we stress that \musketeer is not meant to be used directly.
Rather, \musketeer is a kind of intermediate logic designed for proving the soundness of other light-weight, automatic approaches of ensuring schedule-independent safety such as type systems.
% In other words, \musketeer is not meant to be used by hand on a particular program,
% but rather as, for example, a semantic model of a type system.
%
For instance, our main case study focuses
on using \musketeer to show the soundness of an affine type system guaranteeing schedule-independent safety~(\cref{sec:case_studies}).
Nevertheless, for the sake of explaining the ideas behind \musketeer,
here we explain the reasoning rules
that would allow one to verify manually the schedule-independent safety of $(\exname\,n)$ for all $n$.

\paragraph{Key reasoning rules}
\newcommand{\iscountername}{\textsf{counter}\xspace}
\newcommand{\iscounter}[3]{\iscountername\,#1\,#2\,#3}

\begin{figure}\centering\small
\begin{mathpar}
\inferrule[M-Assert]{}
{\atriple{\iTrue}{\eassert{\val}}{\wal}{\pure{\wal=\vunit \wedge \val=\vtrue}}}

\inferrule[M-KSplit]{}
{\iscounter{\val}{(\qp_1 + \qp_2)}{(i_1 + i_2)} \;\dashv\vdash\; \iscounter{\val}{\qp_1}{\ofs_1} \star \iscounter{\val}{\qp_2}{\ofs_2}}

\inferrule[M-KRef]{}
{\atriple{\iTrue}{\eref\ofs}{\val}{\iscounter{\val}{1}{\ofs}}}

\inferrule[M-KAdd]{}
{\atriple{\iscounter{\val}{\qp}{\ofs}}{\aadd\val{j}}{\_}{\iscounter{\val}{\qp}{(\ofs + j)}}}

\inferrule*[Left=M-KGet]{}
{\atriple{\iscounter{\val}{1}{\ofs}}{\eget\val}{\wal}{\pure{\wal = \ofs} \star \iscounter{\val}{1}{\ofs}}}

\inferrule*[Left=M-Par]
{\triple{\vpre_1}{\expr_1}{\vpost_1} \\ \triple{\vpre_2}{\expr_2}{\vpost_2}}
{\atriplemore{\vpre_1 \star \vpre_2}{\epar{\expr_1}{\expr_2}}{\val}{x}{\exists\val_1\,\val_2\,x_1\,x_2.\;\pure{\val=\vprod{\val_1}{\val_2} \wedge x=(x_1,x_2)} \star \vpost_1\,\val_1\,x_1 \star \vpost_2\,\val_2\,x_2}}
\end{mathpar}
\captionlabel{Reasoning Rules for a Concurrent Counter and Key Reasoning Rules of \musketeer}{fig:keyrules}
\end{figure}

\musketeer takes the form of a unary \SL with triples written
$\triple{\vpre}{\expr}{\vpost}$, where $\vpre$ is a precondition,
$\expr$ the program being verified and $\vpost$ the postcondition.
The postcondition $Q$ is of the form $\lambda\val\,x.\,R$,
where~$\val$ is the value being returned by
the execution of $\expr$ and $x$ is
a \emph{ghost return value}.
We explain ghost return values in detail later, but
% This ghost return value is used to circumvent a limitiation
% of existential quantification in \musketeer;
for now, they can be thought of as a special way to existentially quantify
variables in the postcondition.
This \musketeer triple
guarantees the following hyper-property:
``if one execution of $\expr$ is safe starting from a heap satisfying
$\vpre$ and terminates,
then every execution of $\expr$ is safe starting from a heap satisfying~$\vpre$
and all terminating executions will end in a heap satisfying $\vpost$''.
The upper part of \Cref{fig:keyrules} shows
the main reasoning rules we use for our example
(in these rules, the horizontal bar is an implication in the meta-logic).
% actually, its in Iris
%
While the assertions and rules of \musketeer are similar to standard \SL rules, there are two key differences.
First, \musketeer \emph{does not} provide the usual disjunction or existential elimination rules from \SL.
That is, to prove a triple of the form $\triple{\vpre_1 \vee \vpre_2}{\expr}{\vpost}$, we \emph{cannot} in general do case analysis on the precondition and reduce this to proving $\triple{\vpre_1}{\expr}{\vpost}$ and $\triple{\vpre_2}{\expr}{\vpost}$.
As we will see later, this restriction is necessary because the imprecision in disjunctions and existentials can encode nondeterministic behavior, where different executions pick different witnesses.

Second, unlike traditional \SL rules, rules in \musketeer \emph{do not guarantee safety}.
Rather, they guarantee that \emph{safety is independent of scheduling}.
Thus, these rules often have weaker preconditions than standard \SL rules.
The rule \RULE{M-Assert} illustrates this unusual aspect of \musketeer.
This rule applies to an expression $\eassert\val$, for an arbitrary value $\val$,
and has a \emph{trivial} precondition.
The postcondition has the pure facts that the return value $\wal$
is $\vunit$
and that~$\val$ equals $\vtrue$, \ie that the assert did not fail.
In contrast, the standard \SL rule for $\eassert\val$ requires
the user to \emph{prove} that $\val = \vtrue$!
This is because the expression $\eassert\val$ is safe only if the value $\val=\vtrue$~(\cref{sec:semantics}).
So in conventional \SL, where a triple implies safety, the obligation is to show that the assert will be safe.
However, in \musketeer, the rule \RULE{M-Assert} corresponds exactly to the ``motto'' of \musketeer triples:
 if one execution of $\eassert\val$ is safe
and terminates with value $\wal$
such that $\wal=\vunit$ and $\val=\vtrue$,
then every execution of $\eassert\val$ is safe and terminates with value $\wal=\vunit$, and $\val=\vtrue$ in those executions too.
This property is true in a trivial way:
since the argument $\val$ in $\eassert\val$ is already a value, there is only one possible safe execution
for $\eassert\val$, and such an execution is possible only if $\val=\vtrue$.%
\footnote{Note that \musketeer supports a bind rule
(\RULE{M-Bind}, \cref{fig:almost}) that allows
the user to reason under an evaluation context.
Hence, \musketeer supports reasoning on an expression $(\eassert\expr)$,
for an arbitrary expression $\expr$.
To conduct such a proof,
the user should first apply \RULE{M-Bind} and focus on $\expr$,
show that $\expr$ itself satisfies schedule-independent safety,
and then for any value $\val$ to which $\expr$
may reduce, apply \RULE{M-Assert} on the remaining expression $(\eassert\val)$.}

On the contrary, \RULE{M-Par} has a standard shape.
This rule allows for verifying
the parallel primitive $\epar{\expr_1}{\expr_2}$.
It requires the user to split the precondition
into two parts $\vpre_1$ and $\vpre_2$,
and to establish the two triples
$\triple{\vpre_1}{\expr_1}{\vpost_1}$ and $\triple{\vpre_2}{\expr_2}{\vpost_2}$.
The postcondition of the rule asserts that
the value~$\val$ being returned is an immutable pair $(\val_1,\val_2)$
and the ghost return value~$x$ is itself a pair of two ghost return
values $x_1$ and $x_2$, such that $\vpost\,\val_1\,x_1$ and $\vpost\,\val_2\,x_2$ hold.

\paragraph{Verifying dumas}
The other rules in \Cref{fig:keyrules} are the reasoning rules
for the concurrent counter we use in $\exname$.
They make use of a predicate $\iscounter{\val}{\qp}{i}$,
asserting that $\val$ is a
concurrent counter with fractional ownership $\qp \in (0;1]$.
When $\qp=1$ the assertion represents exclusive ownership of the counter, in which case $i$ is the value stored in the counter.
Otherwise, it asserts ownership of a partial \emph{share} of the counter, and $i$ is the contribution added to the counter with this share.
\RULE{M-KSplit} shows that $\iscountername$ can be split into several shares.
\RULE{M-KRef} verifies $\eref{\ofs}$,
has a trivial precondition and returns a counter initialized to $\ofs$ with fraction $1$.
\RULE{M-KAdd} verifies $\aadd{\val}{j}$, where the share may have an arbitrary fraction.
\RULE{M-KGet} verifies $\eget\val$, requiring that $\iscounter{\val}{1}{i}$ holds.
The fraction is $1$, preventing a concurrent add to $\val$.
Such a concurrent add would introduce nondeterminism based on the relative ordering of the add and get, thereby breaking schedule-independent safety.

Using the above rules, we can show that for any $n$, $\atriple{\iTrue}{(\exname\, n)}{\_}{\iTrue}$, that is, without precondition, the safety of $(\exname\, n)$ is scheduling independent.
To do so, we use \RULE{M-KRef} to initialize the counter, getting $\iscounter{r}{1}{0}$, which
we split into $\iscounter{r}{(1/2)}{0} \star \iscounter{r}{(1/2)}{0}$, and then use \RULE{M-Par}.
The $\iscounter{r}{(1/2)}{0}$ given to each thread is sufficient to reason about the add they each perform,
and when we combine the shares they give back, we get $\iscounter{r}{1}{1844}$.
Using \RULE{M-KRef}, we know that the $\eget{r}$ returns 1844, leaving us to show $\atriple{\iTrue}{\eassert(1844 == n)}{\_}{\iTrue}$.

At this point, we would get stuck in a standard separation logic proof, because the standard rule for assert would require us to prove that $(1844 == n)$ evaluates to $\vtrue$. However, that would only be the case if $n$ was in fact $1844$.
Instead, in \musketeer, we can use a rule showing that $(1844 == n)$ will evaluate to some Boolean $b$, regardless of what value $n$ is.
At that point, we can use \RULE{M-Assert} to conclude, even though we don't know which value $b$ will take.

\subsection{Verifying That One Interleaving is Safe and Terminates with Angelic}
\label{sec:key:angelic}
\begin{figure}\morespacingaroundstar
\[\begin{array}{r@{\;\;\vdash\;\;}l@{\quad}c}
\iTrue & \runex{(\eassert{\vtrue})}{\val}{\pure{\val=\vunit}} & \TirNameStyle{A-Assert}\\
\runex{\expr_1}{\val_1}{\runex{\expr_2}{\val_2}{\post\,(\val_1,\val_2)}} & \run{(\epar{\expr_1}{\expr_2})}\post& \TirNameStyle{A-ParSeqL}\\
\runex{\expr_2}{\val_2}{\runex{\expr_1}{\val_1}{\post\,(\val_1,\val_2)}} & \run{(\epar{\expr_1}{\expr_2})}\post& \TirNameStyle{A-ParSeqR}\\
\iTrue & \runex{(\eref{\ofs})}{\val}{\exists\loc.\, \pure{\val=\loc} \star \loc \mapsto \ofs} & \TirNameStyle{A-Ref}\\
\loc \mapsto \ofs & \runex{(\aadd{\val}{j})}{\_}{\loc \mapsto (\ofs+j)} & \TirNameStyle{A-Add}\\
\loc \mapsto \ofs & \runex{(\eget{\loc})}{\val}{\pure{\val=\ofs} \star \loc \mapsto \ofs} & \TirNameStyle{A-Get}\\
\end{array}\]
\captionlabel{Reasoning Rules for a Concurrent Counter and Key Reasoning Rules of Angelic}{fig:key:angelic}
\end{figure}

Now that we know that for all $n$, $(\exname\, n)$ satisfies schedule-independent safety,
we can prove that $(\exname\, 1844)$ is safe just by showing that \emph{one} interleaving is safe and terminates.
For such a simple example, it would suffice at this point to simply execute $(\exname\, 1844)$ once and observe one safe, terminating execution.
We would then be able to conclude that all possible executions are safe.
However, for more complex examples
(for example, programs that are parameterized by an argument from an infinite type),
we propose \angelic,
a program logic for verifying that one interleaving is safe and terminates.

\angelic uses a form of weakest-precondition reasoning, with specifications
taking the form $\pre \vdash \run{\expr}{\post}$,
where $\pre$ is the precondition, $\expr$ the program being verified,
and $\post$ the postcondition, of the form $\lambda\val.\,\pre'$,
where $\val$ is the value being returned.
In order to guarantee termination,
\angelic's WP is defined as a \emph{total weakest precondition}, that is,
the WP is defined as a least fixpoint and does not mention
the so-called later modality. Such a construction is standard,
\citet[\S4]{krebbers-et-al-25} describes the differences between a WP
for partial and total correctness.
Hence,
$\runex{\expr}{\_}{\iTrue}$
guarantees that one execution of $\expr$ is safe and terminates.

\Cref{fig:key:angelic} presents a few reasoning rules for \angelic.
It is helpful to read these rules backwards, applying the rule to a goal that matches the
right side of the turnstile $\vdash$ and ending up with a goal of proving the left side.
\RULE{A-Assert} verifies an assertion,
for which the argument must be the Boolean true.%
\footnote{\angelic supports a bind rule (\RULE{A-Bind}, \cref{fig:angelic}).
As in \musketeer, \RULE{A-Bind} allows for reasoning under an evaluation context.
In combination with \RULE{A-Assert},
the user may reason about an expression $(\eassert\expr)$
for an arbitrary expression $\expr$.}
Indeed, since \angelic guarantees safety, the proof burden is now to show that the assert will succeed.
\RULE{A-ParSeqL} says that to verify $\epar{\expr_1}{\expr_2}$,
it suffices to verify sequentially $\expr_1$ and then $\expr_2$.
\RULE{A-ParSeqR} lets us verify the reverse order instead, reasoning first about $\expr_2$ and then $\expr_1$.
As we will explain later on~(\cref{sec:angelic_rules}), \angelic more generally allows for selecting
\emph{any} interleaving of steps within $\expr_1$ and $\expr_2$ by ``jumping'' between the two expressions during a proof.
Finally, \RULE{A-Ref}, \RULE{A-Add} and \RULE{A-Get}
shows how to reason on a concurrent counter.
First, these rules do not involve any new predicate, and manipulate
the plain points-to assertion linked with the counter.
Second, no fractions or invariants are involved. Indeed, in \angelic,
there is no need to split and join assertions, as the parallel
primitive can be verified sequentially in any order.

Using these rules,
we can verify that $\vdash \runex{(\exname\, 1844)}{\_}{\iTrue}$ holds,
which implies that there exists one interleaving
that is safe and terminates.
Combined with the fact that this program has schedule-independent safety,
we conclude that $(\exname\, 1844)$ is always safe.

\section{Syntax and Semantics}
\label{sec:syntax_and_semantics}
\lang is a call-by-value lambda calculus
with mutable state and parallelism.
We first present its syntax~(\cref{sec:syntax})
and then its semantics~(\cref{sec:semantics}).
\lang is similar to \heaplang, the language
that ships with Iris, except that it implements structured parallelism
instead of fork-based concurrency.

\subsection{Syntax}
\label{sec:syntax}
\begin{figure}\centering\small
\newcommand{\commentary}[1]{ & \text{\small\it #1} \\}
\newcommand{\defineq}{\mathrel{::=}}
\[
\begin{array}{l@{\quad}r@{\;\defineq\;}l}
\text{Values}\;\Values &\val & \vunit \mid \vbool \in \{\vtrue,\vfalse\} \mid \vint \in \mathbb{Z} \mid \loc \in \Loc \mid \vprod{\val}{\val} \mid \vfun{f}{x}{\expr} \\

\text{Primitives} & \primitive  & + \mid - \mid \times \mid \div \mid\textsf{mod} \mid \symbeq \mid \mathord{<} \mid \mathord{\leq} \mid \mathord{>} \mid \mathord{\geq} \mid \symbor \mid \symband \\

\text{Expressions} &\expr&
\begin{array}[t]{@{}l@{\hspace{8mm}}l@{}}
\begin{array}[t]{@{}ll@{}}
\val,\wal
\commentary{value}
\var
\commentary{variable}
\elet{\var}\expr\expr
\commentary{sequencing}
\eif\expr\expr\expr
\commentary{conditional}
\efun{f}{x}\expr
\commentary{abstraction}
\eapp\expr{\expr}
\commentary{call}
\ecallprim\expr\expr
\commentary{primitive operation}
\eprod\expr\expr
\commentary{product}
\eproj{k \in \{1,2\}}{\expr}
\commentary{projections}
\end{array} &
\begin{array}[t]{@{}ll@{}}
\eassert{\expr}
\commentary{assertion}
\ealloc\expr
\commentary{array allocation}
\eload\expr\expr
\commentary{array load}
\estore\expr\expr\expr
\commentary{array store}
\elength\expr
\commentary{array length}
\epar\expr\expr
\commentary{parallelism}
\erunpar\expr\expr
\commentary{active parallel tuple}
\ecas\expr\expr\expr\expr
\commentary{compare-and-swap}
\end{array}
\end{array}\\
\text{Contexts} & \ectx &
\begin{array}[t]{@{}lllll}
\elet{\var}\khole\expr &\mid \eif\khole\expr\expr &\mid
\ealloc\khole &\mid \elength\khole &\mid \eassert{\khole}\\
\mid \eload\expr\khole &\mid \eload\khole\val &\mid \estore\expr\expr\khole &\mid \estore\expr\khole\val &\mid
\estore\khole\val\val \\
\mid \ecallprim\expr\khole &\mid \ecallprim\khole\val &\mid \eapp\expr{\khole} &\mid \eapp\khole{\val} \\
\mid \ecas\expr\expr\expr\khole &\mid \ecas\expr\expr\khole\val &\mid \ecas\expr\khole\val\val&\mid \ecas\khole\val\val\val\\
\mid \eprod{\expr}{\khole} &\mid \eprod{\khole}{\val} &\mid \eproj{k}{\khole}
\end{array}
\end{array}\]
\captionlabel{Syntax of \lang}{fig:syntax}
%\vspace{-0.5em}
\end{figure}

\cref{fig:syntax} presents the syntax of \lang.
A value~$\val \in \Values$ is either the unit value~$\vunit$, a Boolean~$\vbool \in \{\vtrue,\vfalse\}$,
an idealized integer~$\vint \in \mathbb{Z}$,
a location~$\loc$ from an infinite set of locations $\Loc$,
an immutable product~$\vprod{\val_1}{\val_2}$ of two values,
or a recursive function~$\vfun{f}{x}{\expr}$.

An expression~$\expr$ describe a computation
in \lang.
Recursive functions are written~$\efun{f}{x}{\expr}$.
For non-recursive functions, we write~$\efunnonrec{x}{\expr} \eqdef \efun{\_}{x}{\expr}$.
We define functions with multiple arguments as a chain of function constructors.
Mutable state is available through arrays.
Parallelism is available through a primitive $\epar{\expr_1}{\expr_2}$,
which evaluates to an \emph{active parallel tuple} $\erunpar{\expr_1}{\expr_2}$.
Such a tuple evaluates the two expressions in parallel and returns
their result as an immutable product.
\lang also has a primitive compare-and-swap instruction $\ecas{\expr_1}{\expr_2}{\expr_3}{\expr_4}$,
which targets an array entry and has 4 parameters:
the array location, the offset into the array, the old value and the new~value.
References are defined as arrays of size 1 with the following operations:
\[\begin{array}{c@{\qquad}c@{\qquad}c}
\textsf{ref} \eqdef \efunnonrec{x}{\elet{r}{\ealloc{1}}{\estore{r}{0}{x};\,r}}
 & \textsf{get} \eqdef \efunnonrec{r}{\eload{r}{0}}
 & \textsf{set} \eqdef \efunnonrec{r\,v}{\estore{r}{0}{v}}
\end{array}\]

An evaluation context~$\ectx$ describes an expression with a hole~$\khole$
and dictates the right-to-left evaluation order of \lang.

\subsection{Semantics}
\label{sec:semantics}

%\paragraph{Head Reduction Relation}
\begin{figure}\centering\small
\begin{mathpar}
\inferrule[HeadIfTrue]{}
{\headstep{\eif{\vtrue}{\expr_1}{\expr_2}}\store{\expr_1}\store}

\inferrule[HeadIfFalse]{}
{\headstep{\eif{\vfalse}{\expr_1}{\expr_2}}\store{\expr_2}\store}

\inferrule[HeadCallPrim]
{\purestep{\ecallprim{\val_1}{\val_2}}\val}
{\headstep{\ecallprim{\val_1}{\val_2}}{\store}{\val}{\store}}

\inferrule[HeadAbs]{}
{\headstep{\efun{f}{x}{\expr}}{\store}{\vfun{f}{x}{\expr}}{\store}}

\inferrule[HeadLetVal]{}
{\headstep{\elet{x}{\val}{\expr}}\store{\subst{x}\val{\expr}}\store}

\inferrule[HeadAlloc]
{0 \leq \vint \\ \loc \notin \dom\store}
{\headstep{\ealloc{\vint}}{\store}{\loc}{\blockupd\store{\loc}{\blockrepeat{\vint}{\vunit}}}}

\inferrule[HeadLoad]
{\store(\loc) = \wals \\ 0 \leq \vint < \length\wals \\\\ \wals(i) = \val}
{\headstep{\eload{\loc}{\vint}}{\store}{\val}{\store}}

\inferrule[HeadStore]
{\store(\loc) = \wals \\ 0 \leq \vint < \length\wals}
{\headstep{\estore{\loc}{\vint}{\val}}\store\vunit{\blockupd{\store}{\loc}{\blockupd\wals{\vint}{\val}}}}

\inferrule[HeadAssert]{}
{\headstep{\eassert\vtrue}{\store}{\vunit}{\store}}

\inferrule[HeadProduct]{}
{\headstep{\eprod{\val_1}{\val_2}}{\store}{\vprod{\val_1}{\val_2}}{\store}}

\inferrule[HeadProj]{k \in \{1;2\}}
{\headstep{\eproj{k}{\vprod{\val_1}{\val_2}}}\store{\val_k}\store}

\inferrule[HeadLength]
{\store(\loc) = \wals \\ \vint = \length\wals}
{\headstep{\elength{\loc}}{\store}{\vint}{\store}}

\inferrule[HeadCASSucc]
{ \store(\loc) = \wals \\
  0 \leq \vint < \length\wals \\
  \wals(\vint) = \val
}
{\headstep{\ecas\loc{\vint}\val{\val'}}\store\vtrue{\blockupd\store\loc{\blockupd\wals\vint{\val'}}}}

\inferrule[HeadCASFail]
{ \store(\loc) = \wals \\
  0 \leq \vint < \length\wals \\
  \wals(\vint) = \val_0 \\
  \val_0 \neq \val
}
{\headstep{\ecas\loc{\vint}\val{\val'}}\store\vfalse\store}

\inferrule[HeadCall]{}
{\headstep{\ecall{(\vfun{f}{x}{\expr})}{\val}}{\store}{\subst{f}{(\vfun{f}{x}{\expr})}{\subst{v}{x}\expr}}{\store}}

\inferrule[HeadFork]{}
{\headstep{\epar{\expr_1}{\expr_2}}{\store}{\erunpar{\expr_1}{\expr_2}}{\store}}

\inferrule[HeadJoin]{}
{\headstep{\erunpar{\val_1}{\val_2}}{\store}{\vprod{\val_1}{\val_2}}{\store}}
\end{mathpar}
\captionlabel{Head Reduction Relation}{fig:headsemantics}
%\vspace{-0.5em}
\end{figure}

\Cref{fig:headsemantics} presents the head reduction
relation~$\headstep{\expr}{\store}{\expr'}{\store'}$,
describing a single step of expression~$\expr$ with store
$\store$ into expression~$\expr'$ and store~$\store'$.
A store is a map from location to arrays,
modeled as a list of values.
We write $\emptyset$ for the empty store and $\store(\loc)$ for the list of values at location $\loc$ in~$\store$.
To insert or update a location~$\loc$ with array $\vals$ in store~$\store$,
we write $\blockupd{\store}{\loc}{\vals}$, and similarly write $\blockupd{\vals}{\vint}{\wal}$ to update
offset $\vint$ with value $\wal$ in array $\vals$.
The length of an array $\vals$ is written as $\length\vals$, and $\blockrepeat{\vint}{\val}$
represents an array of size $\vint$ initialized with value $\val$.

Most of the reduction rules are standard.
For example, \RULE{HeadAlloc} allocates an array initialized with the unit value and returns its location, which is selected nondeterministically.
\RULE{HeadLoad} and \RULE{HeadStore} perform loads and stores, respectively.
% \RULE{HeadLength} returns the length of an array.
\RULE{HeadCASSucc} and \RULE{HeadCASFail} performs an atomic compare-and-swap
at an offset in an array.
\RULE{HeadAssert} reduces an assert statement to a unit if the asserted value is $\vtrue$;
asserts of $\vfalse$ are stuck expressions.
\RULE{HeadFork} performs a fork, converting a
primitive par operation into an active parallel tuple.
\RULE{HeadJoin} takes an active
parallel tuple where both sides have reached a value and converts it into an immutable product.

%\paragraph{Main Reduction Relation}
\begin{figure}\centering\small
\newlength\semspace
\setlength\semspace{1.4em}
\begin{mathpar}
\inferrule[StepHead]
{\headstep{\expr}{\store}{\expr'}{\store'}}
{\step{\expr}{\store}{\expr'}{\store'}}
\kern\semspace
\inferrule[StepCtx]
{\step{\expr}{\store}{\expr'}{\store'}}
{\step{\efillctx\ectx\expr}{\store}{\efillctx\ectx{\expr'}}{\store'}}
\kern\semspace
\inferrule[StepParL]
{\step{\expr_1}{\store}{\expr_1'}{\store'}}
{\step{\erunpar{\expr_1}{\expr_2}}{\store}{\erunpar{\expr_1'}{\expr_2}}{\store'}}
\kern\semspace
\inferrule[StepParR]
{\step{\expr_2}{\store}{\expr_2'}{\store'}}
{\step{\erunpar{\expr_1}{\expr_2}}{\store}{\erunpar{\expr_1}{\expr_2'}}{\store'}}
\end{mathpar}
\captionlabel{Main Reduction Relation}{fig:semantics}
%\vspace{-0.5em}
\end{figure}

\Cref{fig:semantics} presents the main reduction
relation~$\step{\expr}{\store}{\expr'}{\store'}$,
describing a parallel step of computation, potentially under an evaluation context.
\RULE{StepHead} performs a head step.
\RULE{StepCtx} performs a computation step under an evaluation
context.
\RULE{StepParL} and \RULE{StepParR} implement parallelism:
these two rules allow for the main reduction relation to perform
nondeterministically a step to the left or right side of an active parallel tuple,
respectively.

We write the reflexive-transitive closure
of the reduction relation as $\steprtc{\expr}{\store}{\expr'}{\store'}$.

\section{A Separation Logic for Proving Schedule-Independent Safety}
\label{sec:logic}
In this section, we present \musketeer in more detail.
%for proving that safety is independent of scheduling.
First, we define schedule-independent safety~(\cref{sec:schedindsafe}).
Next, we introduce our notations for triples and assertions~(\cref{sec:assertions}) and
then present the reasoning rules of \musketeer~(\cref{sec:rules_musketeer}).
We conclude with one of the main technical challenges in working with \musketeer,
the absence of a rule for eliminating existentials,
and explain how we overcame this with the novel concept of \emph{ghost return values}~(\cref{sec:exists}).

\subsection{Definition of Schedule-Independent Safety}
\label{sec:schedindsafe}
\newcommand{\reduciblename}{\textsf{Red}\xspace}
\newcommand{\reducible}[2]{\reduciblename\,#1\,#2}

\newcommand{\notstuckname}{\textsf{Notstuck}\xspace}
\newcommand{\notstuck}[2]{\notstuckname\,#1\,#2}

\newcommand{\safename}{\textsf{Safe}\xspace}
\newcommand{\safe}[1]{\safename\,#1}

\newcommand{\sisname}{\textsf{SISafety}\,\xspace}
\newcommand{\sis}[1]{\sisname\,#1}

\begin{figure}\centering\small
\begin{mathpar}
\inferrule[RedHead]
{\headstep{\expr}{\store}{\expr'}{\store'}}
{\reducible{\expr}{\store}}

\inferrule[RedCtx]
{\reducible{\expr}{\store}}
{\reducible{(\efillctx\ectx\expr)}{\store}}

\inferrule[RedPar]
{ \expr_1 \notin \Values \,\lor\, \expr_2 \notin \Values \\\\
\expr_1 \notin \Values \implies  \reducible{\expr_1}{\store} \\
\expr_2 \notin \Values \implies  \reducible{\expr_2}{\store}}
{\reducible{(\erunpar{\expr_1}{\expr_2})}{\store}}
\end{mathpar}

\[\begin{array}{r@{\;\;\eqdef\;\;}l}
\notstuck{\expr}{\store} & \expr \in \Values \;\lor\; \reducible{\expr}{\store}\\
\safe{\expr} & \forall \expr'\,\store'.\;(\steprtc{\expr}{\emptyset}{\expr'}{\store'}) \implies \notstuck{\expr'}{\store'}\\
\sis\expr & \forall \val\,\sigma.\;(\steprtc{\expr}{\emptyset}{\val}{\store}) \implies \safe{\expr}
\end{array}\]
\captionlabel{Definition of the \reduciblename, \notstuckname, \safename, and \sisname Predicates}{fig:red}
\end{figure}

% Before diving into the details of \musketeer,
Let us make formal the
definition of \emph{schedule-independent safety},
that is, the property guaranteeing our motto ``\motto{\expr}{\expr}''.

What does it mean for a parallel program to be safe?
We say that the configuration $\config{\expr}{\store}$ is \emph{not stuck}
if either $\expr$ is a value,
or every parallel task in $\expr$ that has not reached a value can
take a~step---in the latter case, we call the configuration \emph{reducible}.
A program is defined to be safe if every configuration it can reach is not stuck.
In particular, if a program $\expr$ is safe, then no assertion in $\expr$ can fail, since an assert of a $\vfalse$ value is not reducible.

\Cref{fig:red} gives the formal definitions.
The upper part of \Cref{fig:red} defines the property $\reducible{\expr}{\store}$,
asserting that the configuration $\config{\expr}{\store}$ is reducible.
\RULE{RedHead} asserts that if $\expr$ can take a head step, then it is reducible.
\RULE{RedCtx} asserts that the reducibility of an expression $\efillctx{\ectx}{\expr}$
follows from reducibility of $\expr$.
\RULE{RedPar} asserts that an active parallel tuple $\erunpar{\expr_1}{\expr_2}$
is reducible if at least one sub-expression is not a value (otherwise, a join is possible),
and each sub-expression that is not a value is reducible.
The lower part of \Cref{fig:red}
asserts that the property $\notstuck{\expr}{\store}$ holds if and only
if either $\expr$ is a value or $\reducible{\expr}{\store}$ holds.
Then, $\safe{\expr}$ says that if $\config{\expr}{\emptyset}$ can reach $\config{\expr'}{\store'}$ in zero or more steps, then $\notstuck{\expr'}{\store'}$.
Finally, the main property $\sis{\expr}$, asserting that the safety of $\expr$ is schedule-independent, is defined.
The property says that if some execution of $\expr$ reaches a value~$\val$, then $\expr$ is safe.
% and that some other execution of $\expr$, potentially following
% a different interleaving, reaches a state~$\config{\expr'}{\store'}$.
% The property concludes that $\notstuck{\expr'}{\store'}$ holds.
The soundness \cref{thm:musketeer} of \musketeer guarantees that, for
a verified program~$\expr$, the property $\sis\expr$ holds.

\subsection{Triples and Assertions}
\label{sec:assertions}
As we saw, \musketeer is a \SL whose main judgement takes the
form of a triple $\triple{\vpre}{\expr}{\vpost}$.
In this triple, $\vpre$ is the \emph{precondition},
$\expr$ the program being verified,
and $\vpost$ the \emph{postcondition}.
The postcondition is of the form $\lambda\val\,x.\,\vpre'$,
where~$\val$ is the value being returned by
the execution of $\expr$ and $x$ is
a \emph{ghost return value} returned by the verification of $\expr$.
% We discuss the need for ghost return values later~(\cref{sec:exists}).
% For now, a ghost return value can be understood
% as an existentially-quantified variable.
%
Both $\vpre$ and $\vpre'$ are \SL assertions,
and can be understood as \emph{heap predicates}:
they describe the content of a heap.
We write $\vpre \star \vpre'$ for the separating conjunction,
$\vpre \wand \vpre'$ for the separating implication and $\pure{P}$ when the property $P$
holds in the meta-logic (\ie Rocq).
\musketeer offers fractional~\citep{bornat-permission-accounting-05, boyland-fractions-03} points-to assertions
$\loc \fpointsto{\qp} \vals$.
This assertion says that the location $\loc$ points to the array $\vals$ with fraction $\qp \in (0;1]$.
When $\qp = 1$ we simply write $\loc \pointsto \vals$.
We use the term $\vProp$ for the type of assertions that can be used in \musketeer pre/post-conditions.

As described before, the \musketeer triple $\triple{\vpre}{\expr}{\vpost}$
can be intuitively read as implying the following hyper-property:
``if one execution of $\expr$ is safe starting from a heap satisfying
$\vpre$ and terminates,
then every execution of $\expr$ is safe starting from a heap satisfying~$\vpre$
and all terminating executions will end in a heap satisfying $\vpost$''.
If $\vpre$ and $\vpost$ are trivial, then this implies the $\sisname$
property.
This is captured formally in the soundness theorem of the logic.
\begin{theorem}[Soundness of \musketeer]\label{thm:musketeer}
If $\atriple\iTrue\expr{\_}{\iTrue}$ holds,
then $\sis\expr$ holds.
\end{theorem}

At first, this soundness theorem might seem weak, since it focuses on \emph{safety} of all executions.
What if we instead want to show that every terminating execution satisfies a stronger postcondition~$Q$?
In general, \cref{thm:musketeer} does not directly imply such a stronger property, but recall that in \lang, safety implies that no assert fails.
Thus, by annotating a program with appropriate assert statements,
we can encode various specifications in terms of safety.
We illustrated this aspect in our $\exname$ example~(\cref{sec:key:example}),
where safety implied that the return value across all executions would equal a particular number.

Although \musketeer is a unary logic with judgements referring to a single program $\expr$,
the above statement reveals that the judgements are relating together
multiple executions of that program.
To make this work, under the hood, \musketeer's $\vProp$ assertions
describe not one but \emph{two} heaps,
corresponding to two executions of the program.
This has ramifications for some proof rules~(\cref{sec:exists}).
Later, we will see how $\vProp$ assertions can be encoded
into assertions in a relational logic that
makes these two different heaps more explicit.
% We refer to the assertions in \musketeer as $\vProp$
% For reasons discussed in \Cref{sec:chained}, we refer
% to these two heaps as the left-hand side and right-hand side heaps.
%

\subsection{Reasoning Rules for Musketeer}
\label{sec:rules_musketeer}
\begin{figure}\centering\small\morespacingaroundstar\morespacingaroundwedge
\begin{mathpar}
% \inferrule[M-Assert]{}
% {\atriple{\iTrue}{\eassert{\val}}{\wal}{\pure{\wal=\vunit \wedge \val=\vtrue}}}
%
% \inferrule[M-LetVal]
% {\triple{\pre}{\subst{x}{\val}{\expr}}{\post}}
% {\triple{\pre}{\elet{x}{\val}{\expr}}{\post}}
%
\inferrule[M-If]
{(\,\val = \vtrue \implies \triple{\vpre}{\expr_1}{\vpost}\,) \\\\
(\,\val = \vfalse \implies \triple{\vpre}{\expr_2}{\vpost}\,)
}
{\triple{\vpre}{\eif{\val}{\expr_1}{\expr_2}}{\vpost}}

\inferrule[M-Conseq]
{\vpre \wand \vpre' \\ \triple{\vpre'}{\expr}{\vpost'} \\
\forall\val\,x.\;\vpost\,\val\,x \wand \vpost'\,\val\,x}
{\triple{\vpre}{\expr}{\vpost}}

\inferrule[M-Val]
{\vpre \wand \vpost\,\val\,x}
{\triple{\vpre}{\val}{\vpost}}

\inferrule*[Left=M-Alloc]{}
{\atriplemore{\iTrue}{\ealloc{\wal}}{\val}{(\loc,\vint)}{\pure{\val=\loc \wedge \wal=\vint \wedge 0 \leq \vint} \star \loc \pointsto \blockrepeat{\vint}{\vunit}}}

\inferrule*[Left=M-Load]{}
{\atriplemore
{\loc \fpointsto{\qp} \vals}
{\eload{\loc}{\wal}}{\val'}\ofs
{\pure{\wal=\ofs \wedge 0 \leq \ofs < \length{\vals} \wedge \vals(\ofs)=\val'} \star \loc \fpointsto{\qp} \vals}}

\inferrule*[Left=M-Store]{}
{\atriplemore{\loc \pointsto \vals}
{\estore{\loc}{\wal}{\val'}}
{\val''}\ofs{\pure{\val''=\vunit \wedge \wal=\ofs \wedge 0 \leq \ofs < \length{\vals}} \star \loc \pointsto {\blockupd\vals\ofs{\val'}}}}
%
% \inferrule*[Left=M-Par]
% {\triple{\vpre_1}{\expr_1}{\vpost_1} \\ \triple{\vpre_2}{\expr_2}{\vpost_2}}
% {\atriplemore{\vpre_1 \star \vpre_2}{\epar{\expr_1}{\expr_2}}{\val}{x}{\exists\val_1\,\val_2\,x_1\,x_2.\;\pure{\val=\vprod{\val_1}{\val_2} \wedge x=(x_1,x_2)} \star \vpost_1\,\val_1\,x_1 \star \vpost_2\,\val_2\,x_2}}

\inferrule[M-Bind]
{\atriplemore{\vpre}{\expr}{\val}{x}{\vpost'\,\val\,x} \\
\forall\val\,x.\;\triple{\vpost'\,\val\,x}{\efillctx{\ectx}{\val}}{\vpost}}
{\triple{\vpre}{\efillctx{\ectx}{\expr}}{\vpost}}

\inferrule[M-Frame]
{\triple{\vpre}{\expr}{\vpost}}
{\atriplemore{\vpre \star \vpre'}{\expr}{\val}{x}{\vpost\,\val\,x \star \vpre'}}
\end{mathpar}
\captionlabel{Selected Reasoning Rules of \musketeer (extends \cref{fig:keyrules})}{fig:almost}
\end{figure}

\Cref{fig:almost} presents selected reasoning rules of \musketeer.
Recall that because \musketeer triples do not imply safety, these rules differ from familiar \SL rules.
We have previously seen this in the rule \RULE{M-Assert}.
A similar phenomenon happens in \RULE{M-If}, which targets the expression $\eif{\val}{\expr_1}{\expr_2}$.
In standard \SL, one must prove that $\val$ is a Boolean, since otherwise the if-statement would get stuck.
However, in \RULE{M-If}, the user does not have to prove that $\val$
is a Boolean.
Instead, the rule requires the user to verify the two
sides of the if-statement under the hypothesis that $\val$ was the
Boolean associated with the branch.

\RULE{M-Alloc}, \RULE{M-Load} and \RULE{M-Store}
are similar to their standard \SL counterparts,
except that they do not require the user to show that
the allocation size or the loaded or stored offset are valid integers.
\RULE{M-Alloc} targets the expression $\ealloc\wal$
and has a trivial pre-condition.
The postcondition asserts that
the value being returned is a location $\loc$ and that $\wal$
is a non-negative integer--recall that we can think of the ghost return value $(\loc, i)$
as if it were just a special way of existentially quantifying the variables $\loc$ and $i$ in the postcondition.
The postcondition additionally contains the points-to assertion
$\loc \pointsto \blockrepeat{\vint}{\vunit}$ asserting that $\loc$
points to the array of size $\vint$ initialized with the unit value.
\RULE{M-Load} and \RULE{M-Store} follow the same pattern.

\RULE{M-Alloc} might surprise the reader, since based on the interpretation of triples we described above, the postcondition seems to imply that \emph{every} execution of the allocation
will return the same location~$\loc$.
Yet allocation in \lang is \emph{not} deterministic.
The resolution of this seeming contradiction, is that because \lang does not allow for ``constructing'' a location
(\eg. transforming an integer into a location),
there is no way for the program to observe the nondeterminism of allocations.
Hence, from the reasoning point-of-view we can conduct the proof \emph{as if}
allocations were made deterministically.
This subtlety will appear in the model of \musketeer~(\cref{sec:chained_rules}).

\RULE{M-Bind} allows for reasoning under a context,
and is very similar to the standard \SL \textsc{Bind} rule, except that in the second premise, we
quantify over not just the possible return values $\val$, but also the ghost return value $x$.
% \todo{perhaps more on the ghost return value}
%
\RULE{M-Val} allows for concluding a proof about a value, allowing the user of the rule to pick an arbitrary
ghost return value $x$.
\RULE{M-Frame} shows that \musketeer supports framing.
\RULE{M-Conseq}
is the consequence rule of \musketeer: it allows for weakening
the precondition and strengthening the postcondition.

\subsection{Existential Reasoning with Ghost Return Values}
\label{sec:exists}
\bgroup\morespacingaroundstar

In \SL, existential quantification is essential for modularity.
Among other things, it allows for concealing
intermediate pointers behind an abstraction barrier.
To see how this is typically done, let us consider an example making use of the following $\indirection$ function that creates a reference to a reference:
\[\indirection \eqdef \efunnonrec{v}{\eref{(\eref{\val})}}\]
Without using ghost return value, a possible specification for $\indirection\,\val$ would be:
\[\atriple{\iTrue}{\indirection\,\val}{\wal}
{\exists\loc.\,\pure{\wal=\loc} \star \exists \loc'.\, \loc \pointsto [\loc'] \star \loc' \pointsto [\val]}\]
In the above specification, the first existential quantification on $\loc$
does not hide or abstract over anything, since the returned value~$\wal$ uniquely characterizes $\loc$.
However, the existential quantification on $\loc'$ is more interesting,
as it forms an abstraction barrier: it hides this intermediate location.
Let us now focus on a client of $\indirection$,
and try to verify the following triple:
\[\atriple{\iTrue}{\eget{(\indirection\,\val)}}{\_}{\iTrue}\]
Making use of \RULE{M-Bind} and then applying the above specification for $\indirection$, we obtain:
\[\atriple{\exists\loc.\,\pure{\wal=\loc} \star \exists \loc'.\, \loc \pointsto [\loc'] \star \loc' \pointsto [\val]}{\eget{\wal}}{\_}{\iTrue} \qquad (\TirNameStyle{intermediate})\]
We now need to \emph{eliminate} the existentials on $\loc$ and $\loc'$
in the precondition by introducing universally-quantified variables in the meta-logic.
More precisely, we would like to apply the following standard \SL rule:
\begin{mathpar}
\inferrule
{\forall x.\;\triple{\vpre\,x}\expr\vpost}
{\triple{\exists x.\,\vpre\,x}{\expr}{\vpost}}
\end{mathpar}
However, \musketeer \emph{does not support this rule}.
Indeed, although \musketeer is formulated as a unary logic,
it relates two executions of the same program.
As we previously alluded to~(\cref{sec:assertions}),
\musketeer's $\vProp$ assertions are,
under the hood, tracking not one, but two heaps: one for each execution of
the same program.
The fact that preconditions describe two heaps
implies that the precondition
$\exists x.\,\vpre\,x$ has two interpretations---one
for each heap of the two executions of $\expr$ being tracked by the triple.
Although the precondition holds in both heaps, the witness~$x$ might differ between the two.
Whereas, in the premise of the above rule, quantifying over~$x$ at the meta-level means that~$x$ is treated
as the same in both executions.
We present a detailed example of such a case in \citeappendix{appendix:counterexample}.

%What are the limitations of \emph{not} having the above rule?
%Most of the time, it is not problematic, because we offer
As a result, \musketeer only supports the weaker rule \RULE{M-ElimExist},
allowing an existential to be eliminated
when the precondition guarantees that the witness is unique.
\begin{mathpar}
\kern2em\inferrule*[Left=M-ElimExist]
{(\forall x.\, \vpre\,x \wand \pure{U\,x}) \\
(\forall x\,y.\, U\,x \wedge U\,y \implies x=y)\\
(\forall x.\;\triple{\vpre\,x}{\expr}{\vpost})}
{\triple{\exists x.\, \vpre\,x}{\expr}{\vpost}}
\kern-2em
\end{mathpar}

For example, in the above \RULE{intermediate} triple,
\RULE{M-ElimExist} would allow to eliminate the quantification on $\loc$, since it is uniquely characterized by $\wal$.
However, \RULE{M-ElimExist} is tedious to use in practice.
Moreover, sometimes objects are \emph{not}
uniquely characterized by the precondition,
and yet are chosen deterministically, so that the witnesses ought to be the same in both executions.
For example, in the above \RULE{intermediate} triple,
\RULE{M-ElimExist} cannot be used to eliminate the quantification on $\loc'$.

To solve this issue, we introduce ghost return values.
In a Musketeer triple $\atriplemore{\vpre}{\expr}{\val}{x}{\vpost\,\val\,x}$,
the ghost return value $x$ is an object (of an arbitrary type, which is formally a parameter of the triple)
that will eventually be chosen by the user when they apply \RULE{M-Val}.
We think of the bound variable~$x$ as if it were existentially quantified, but the key is that the eventual ``witness'' selected when using \RULE{M-Val} will be the same across the two executions under consideration.
As a result, instead of having to use the weak \RULE{M-ElimExist} to eliminate $x$, the ghost return value is \emph{automatically} eliminated in a strong way by \RULE{M-Bind}.

Let us go back to our $\indirection$ example.
We prove a specification for $\indirection$ in which $\loc'$ is bound in a ghost return value, instead of as an existential:
\[\atriplemore{\iTrue}{\indirection\,\val}{\val}{\loc'}{\exists\loc.\,\pure{\val=\loc} \star \loc \pointsto [\loc'] \star \loc' \pointsto [\val]}\]
As we use this specification to reason about $(\eget\ (\indirection\, \val))$,
\RULE{M-Bind}
will eliminate $\loc'$ automatically, and we can use \RULE{M-ElimExist} to eliminate $\loc$,
reducing the proof to:
\[\atriple{\loc \pointsto [\loc'] \star \loc' \pointsto [\val]}{\eget{\loc}}{\_}{\iTrue}\]
allowing us to proceed and conclude, since there is no longer an existential to eliminate.
\egroup

We extensively use ghost return values
for the verification of \minidet, our case study~(\cref{sec:case_studies}).
For instance, we use a ghost return value to record the content
of references in the typing environment.

\section{Unchaining the Reasoning with Chained Triples}
\label{sec:chained}
For an expression $\expr$, a \musketeer triple guarantees
the property ``\motto{\expr}{\expr}''.
% This sentence includes two occurrence of the expression~$\expr$.
In order to justify the validity of the reasoning rules
for \musketeer triples,
we generalize the above property
and define an intermediate
logic called \chainedlog which targets two
expressions $\expr_l$ and $\expr_r$ and guarantees
the property ``\motto{\expr_l}{\expr_r}''.
We first present chained triples~(\cref{sec:chained_start})
and present some associated reasoning rules~(\cref{sec:chained_rules}).
Finally, we explain how we encode \musketeer triples using chained triples~(\cref{sec:musketeer_def}).

\subsection{Chained Triples as a Generalization of Musketeer Triples}
\label{sec:chained_start}

In \chainedlog, a chained triple takes the form:
\[\ctriplebase{\pre_l}{\expr_l}{\post_l}{\pre_r}{\expr_r}{\post_r}\]
The assertions~$\pre_l$ and~$\pre_r$ are the preconditions
of $\expr_l$ and $\expr_r$, respectively.
The assertions~$\post_l$ and~$\post_r$ are both of the form $\lambda\val.\,\pre$,
where $\val$ is a return value, and are the postconditions
of $\expr_l$ and $\expr_r$, respectively.
Intuitively, the above chained triple says that,
if there exists a reduction of $\expr_l$
starting from a heap satisfying $\pre_l$,
that is safe and terminates on
a final heap with a value $\val_l$ satisfying $\post_l\,\val_l$, then
every reduction of $\expr_r$
starting from a heap satisfying $\pre_r$
is safe and if it terminates,
it does so on
a final heap with a value $\val_r$ satisfying $\post_r\,\val_r$.
Moreover, chained triples guarantee determinism (for simplicity, see our commentary of \RULE{C-Par}), that is,
$\val_l=\val_r$.
Formally, we have the following soundness theorem:
\begin{theorem}[Soundness of Chained Triples]\label{thm:chained}
If $\ctriplebase\iTrue{\expr_l}{\_.\,\iTrue}\iTrue{\expr_r}{\lambda\_.\,\iTrue}$
holds,
and if there exists a value $\val$ and a store $\store$
such that $\steprtc{\expr_l}\emptyset\val\store$,
then the property $\safe{\expr_r}$ holds.
\end{theorem}

In particular, chained triples do not guarantee safety for $\expr_l$,
but they do guarantee safety for $\expr_r$.
We call the triples ``chained'' because enjoy the following rule that allows us to chain facts from one execution to the other:
\begin{mathpar}\morespacingaroundstar
\inferrule*[Left=C-Chain]
{\ctriplebase{\pre_l}{\expr_l}{\lambda\val_l.\,\post_l\,\val_l\ast \pre}{\pre_r}{\expr_r}{\lambda\val_r.\,\post_r}}
{\ctriplebase{\pre_l}{\expr_l}{\lambda\val_l.\,\post_l\,\val_l}{\pre \wand \pre_r}{\expr_r}{\lambda\val_r.\,\post_r}}
\end{mathpar}
It is best to read this rule from the bottom up.
Below the line, using the precondition for $\expr_r$ requires showing $\pre$.
Above the line, the rule allows us to discharge this assumption by showing that $\pre$ holds in the postcondition of $\expr_l$.
That is, if some knowledge $\pre$ is needed in order to verify
the safety of $\expr_r$,
then this knowledge can be gained from an execution of $\expr_l$.

% \paragraph{Assertions of Chained Triples}
Assertions $\pre$ of \chainedlog are ground Iris assertions of type \iProp{}.
As previously intuited~(\cref{sec:assertions}),
they include two forms of points-to assertions,
one for each side of the triple.
We write $\loc \fpointstox{\gol}\qp \vals$ the points-to assertion
for the left expression, and $\loc \fpointstox{\gor}\qp \vals$
for the right expression.
Moreover, \chainedlog makes use of a \emph{left-allocation token},
written $\leftalloc\loc$.
This (non-persistent) assertion witnesses that $\loc$ has been
allocated by the left expression and plays a key role for allocations.

\subsection{Reasoning Rules for Chained Triples}
\label{sec:chained_rules}
\begin{figure}\centering\small\morespacingaroundstar
\begin{mathpar}
\inferrule[C-AssertL]
{\val=\vtrue \implies \ctriple{\pre_l}{\ectxs_l}{\vunit}{\post_l}{\pre_r}{\ectxs_r}{\expr_r}{\post_r}}
{\ctriple{\pre_l}{\ectxs_l}{\eassert\val}{\post_l}{\pre_r}{\ectxs_r}{\expr_r}{\post_r}}

\inferrule[C-AssertR]
{\ctriple{\pre_l}{\ectxs_l}{\expr_l}{\post_l}{\pre_r}{\ectxs_r}{\vunit}{\post_r}}
{\ctriple{\pre_l}{\ectxs_l}{\expr_l}{\post_l}{\pre_r}{\ectxs_r}{\eassert{\vtrue}}{\post_r}}

\inferrule*[Left=C-AllocL]
{\forall\loc\,\vint.\;\val=\vint \wedge 0 \leq \vint \implies \ctriple{\pre_l \star \loc \pointstox\gol \blockrepeat{\vint}{\vunit} \star \leftalloc{\loc}}{\ectxs_l}{\loc}{\post_r}{\pre_r}{\ectxs_r}{\expr_r}{\post_r}}
{\ctriple{\pre_l}{\ectxs_l}{\ealloc{\val}}{\post_r}{\pre_r}{\ectxs_r}{\expr_r}{\post_r}}

\inferrule*[Left=C-AllocR]
{0 \leq \vint \\ \ctriple{\pre_l}{\ectxs_l}{\expr_l}{\post_l}{\pre_r \star \loc \pointstox\gor \blockrepeat{\vint}{\vunit} }{\ectxs_r}{\loc}{\post_r}}
{\ctriple{\pre_l}{\ectxs_l}{\expr_l}{\post_l}{\pre_r \star \leftalloc{\loc}}{\ectxs_r}{\ealloc{\vint}}{\post_r}}

\inferrule*[Left=C-LoadL]
{\forall\vint\,\wal.\;\val=\vint \wedge 0 < \vint \leq \length\vals \wedge \wal = \vals(\ofs) \implies \ctriple{\pre_l \star \loc \fpointstox{\gol}{\qp} \vals}{\ectxs_l}{\wals}{\post_r}{\pre_r}{\ectxs_r}{\expr_r}{\post_r}}
{\ctriple{\pre_l\star \loc \fpointstox{\gol}{\qp} \vals}{\ectxs_l}{\eload{\loc}{\val}}{\post_r}{\pre_r}{\ectxs_r}{\expr_r}{\post_r}}

\inferrule*[Left=C-LoadR]
{0 < \vint \leq \length\vals \wedge \wal = \vals(\ofs) \\
\ctriple{\pre_l}{\ectxs_l}{\expr_l}{\post_l}{\pre_r \star \loc \fpointstox{\gor}{\qp} \vals}{\ectxs_r}{\wal}{\post_r}}
{\ctriple{\pre_l}{\ectxs_l}{\expr_l}{\post_l}{\pre_r \star \loc \fpointstox{\gor}{\qp} \vals}{\ectxs_r}{\eload\loc{\vint}}{\post_r}}

\inferrule*[Left=C-Par]
{\ctriplebase{\pre_{l1}}{\expr_{l1}}{\post_{l1}}{\pre_{r1}}{\expr_{r1}}{\post_{r1}}\\
\ctriplebase{\pre_{l2}}{\expr_{l2}}{\post_{l2}}{\pre_{r2}}{\expr_{r2}}{\post_{r2}} \\
\forall \val_1\,x_1\,\val_2\,x_2.\;
\ctriple{\post_{l1}\val_1\,x_1 \star \post_{l2}\,\val_2\,x_2}{\ectxs_l}{\vprod{\val_1}{\val_2}}{\post_l}{\post_{r1}\val_1\,x_1 \star \post_{r2}\,\val_2\,x_2}{\ectxs_r}{\vprod{\val_1}{\val_2}}{\post_r}
}
{\ctriple{\pre_{l1} \star \pre_{l2}}{\ectxs_l}{\epar{\expr_{l1}}{\expr_{l2}}}{\post_l}{\pre_{r1} \star \pre_{r2}}{\ectxs_r}{\epar{\expr_{r1}}{\expr_{r2}}}{\post_r}}

\inferrule*[Left=C-FrameL]
{\ctriplebase{\pre_l}{\expr_l}{\post_l}{\pre_r}{\expr_r}{\post_r}}
{\ctriplebase{\pre_l \star \pre_0}{\expr_l}{\post_l \star \pre_0}{\pre_r}{\expr_r}{\post_r}}

\kern2em
\inferrule*[Left=C-FrameR]
{\ctriplebase{\pre_l}{\expr_l}{\post_l}{\pre_r}{\expr_r}{\post_r}}
{\ctriplebase{\pre_l}{\expr_l}{\post_l}{\pre_r\star \pre_0}{\expr_r}{\lambda\val_r.\,\post_r\,\val_r\star \pre_0}}
\kern-4em
\end{mathpar}
\captionlabel{Selected Reasoning Rules for Chained Triples}{fig:dwp}%
\vspace{-0.1em}
\end{figure}

\Cref{fig:dwp} presents selected reasoning rules for chained triples.
Before commenting on these rules, let us underline
a caveat of chained triples,
explaining in part why we only use them as a model for \musketeer :
chained triples do \emph{not} support a \textsc{Bind} rule.%
\footnote{The absence of a \textsc{Bind}
rule comes from the chaining intention
of these triples:
the user needs to terminate the reasoning on the whole
left-hand side expression before reasoning on the right-hand side.}
Hence, non-structural rules for chained triples explicitly
mentions a stack of contexts, written $\ectxs$.

Let us again start with the rules for reasoning about an assertion.
\RULE{C-AssertL} allows for reasoning about $\eassert{\val}$
on the left-hand side.
Because this rule targets the left hand-side, there is no
safety-related proof obligation, hence the premise of the rule
allows the user to suppose that $\val=\vtrue$.
\RULE{C-AssertR} is, on the contrary, similar to a standard \SL rule for assertions:
the assertion must target a Boolean, and this Boolean must be true.

\RULE{C-AllocL} allows for reasoning about an allocation of an array on the left-hand side.
Again, there is no safety proof obligation, so the user gets
to suppose that the argument of the allocation is a non-negative integer.
The precondition is then augmented with a points-to
assertion to a universally quantified location $\loc$
as well as the allocation token $\leftalloc{\loc}$.
This latter assertion plays a role in \RULE{C-AllocR},
which allows for reasoning about an allocation on the right-hand side.
Indeed, the assertion $\leftalloc{\loc}$ appears in the precondition of
the right-hand side.
This assertion allows for \emph{predicting} the location
allocated on the right-hand side.
As a result, the premise of \RULE{C-AllocR} does not universally
quantify over the location allocated--the name $\loc$ is reused.
The user can transmit a $\leftalloc\loc$ assertion from
the left-hand side to the right-hand side using \RULE{C-FrameL} and \RULE{C-Chain}.

This rule may seem surprising, since allocation is nondeterministic in \lang, yet
this rule appears to ensure that the right-hand side allocation returns the \emph{same} location
as the left-hand side.
The key is that a right-hand points-to assertion of the form $\loc \fpointstox{\gor}\qp \vals$
does \emph{not} mean that the specific location $\loc$ has that value in the right-hand side execution.
Rather, it means that there exists some location which points to $\vals$ on the right-hand side, and we can reason \emph{as if} that location were equivalent to $\loc$, under some implicit permutation renaming of locations.
In other words, as we alluded to earlier in \Cref{sec:rules_musketeer} when discussing the nondeterminism of allocation in \musketeer, the logic ensures that the specific location of an allocation does not matter, since we do not support casting integers to pointers.

Our approach of using the $\leftalloc{\loc}$ assertion
has two consequences.
First, as we will see~(\cref{sec:musketeer_def}),
it will allow us to define \musketeer triples
in terms of chained triples where both the left- and right-hand side
coincide; such a definition would be impossible if
the allocation on the left and on the right-hand side could return different names.
Second, it bounds the number of allocations on the right-hand side
by the number of allocations on the left-hand side.
We posit that this limitation can be lifted by distinguishing
between synchronized locations, whose name come from the left-hand side,
and unsynchronized one.
We were able to conduct our case studies without such a feature.

\RULE{C-LoadL} and \RULE{C-LoadR} follow the same spirit as the
previous rules: the rule for the left-hand side
has no safety proof obligation, but the right-hand size has a standard \SL shape.

\RULE{C-Par} targets a parallel primitive
and is a \emph{synchronization point}:
both the left- and right-hand side
must face a parallel primitive.
The rule mimics a standard \textsc{Par} rule
on both sides at once.
In particular, it requires the user to split
the preconditions of the left- and right-hand sides,
which will be given to the corresponding side of the
active parallel pair.
The bottom premise of \RULE{C-Par}
requires the user to verify the continuation,
after the execution of the parallel primitive ended.
This premise also show the (external) determinism
guaranteed by chained triple: the execution is resumed
on both sides with the same result of the parallel execution:
the immutable pair $(\val_1,\val_2)$.
Note also that both sides agree on the ghost return values.%

\subsection{Encoding \musketeer in \chainedlog}
\label{sec:musketeer_def}
\begin{figure}\centering\small\morespacingaroundstar
\[\begin{array}{r@{\;\;\eqdef\;\;}l@{\quad\qquad}r@{\;\;\eqdef\;\;}l}
\vProp&\mathbb{B} \rightarrow \iProp & \forall x.\, \vpre\,x&\lambda b.\, \forall x.\,\vpre\,x\,b\\
\vpre_1 \star \vpre_2&\lambda b.\, \vpre_1\,b \star \vpre_2\,b & \exists x.\, \vpre\,x&\lambda b.\, \exists x.\,\vpre\,x\,b \\
\vpre_1 \wand \vpre_2&\lambda b.\, \vpre_1\,b \wand \vpre_2\,b &
\loc \fpointsto{\qp} \vals&\lambda b.\, \eif{b}{\loc \fpointstox\gol{\qp} \vals}{\loc \fpointstox\gor{\qp} \vals}\\[0.5em]
\end{array}\]
\vspace{-0.2em}%
\captionlabel{Definition of $\vProp$ assertions}{fig:vprop}%
\vspace{-0.2em}
\end{figure}

\begin{figure}\centering\small\morespacingaroundstar
\begin{align*}
&\triple{\vpre}{\expr}{\vpost} \;\eqdef\; \forall \ectxs\,\pre_l\,\pre_r\,\post_l\,\post_r.\\
&\qquad \big(\,\forall\val\,x.\,\ctriple{\vpost\,\val\,x\,\vtrue \star \pre_l}{\ectxs}{\val}{\post_l}{\vpost\,\val\,x\,\vfalse \star \pre_r}{\ectxs}{\val}{\post_r} \,\big) \wand\\
&\qquad \ctriple{\vpre\,\vtrue \star \pre_l}{\ectxs}{\expr}{\post_l}{\vpre\,\vfalse\star \pre_r}{\ectxs}{\expr}{\post_r}
\end{align*}
\vspace{-0.2em}%
\captionlabel{Definition of \musketeer Triples}{fig:tripledef}%
\vspace{-0.2em}
\end{figure}

We now discuss how to encode \musketeer into \chainedlog.
Recall that \musketeer's assertions have the type $\vProp$.
We encode these as functions from Booleans to $\iProp$,
the ground type of \chainedlog assertions.
The idea is that the $\vProp$ tracks two heaps, and we use the Boolean parameter of the function to indicate which
side of the \chainedlog the assertion is
being interpreted to: $\textsf{true}$ indicates the left side, and $\textsf{false}$ the right side.
The formal definition of $\vProp$ assertions appears in
\Cref{fig:vprop}.
The Boolean parameter is threaded through
the separating star and implication, and similarly
for the $\forall$ and $\exists$ quantifier.
The points-to assertion simply cases over the Boolean and returns the
left or right version of the points-to.
Entailment is defined as $P_1 \vdash P_2 \,\eqdef\, \forall b.\,P_1\,b \vdash P_2\,b$.

Next, we can encode \musketeer triples as shown in \Cref{fig:tripledef}.
A \musketeer triple $\triple{\vpre}{\expr}{\vpost}$
is mapped to a chained triple where both sides refer to the expression $\expr$ use the precondition $\vpre$ instantiated with Booleans corresponding to the appropriate side.
Because chained triples do not support a bind rule, the encoding is written in a continuation passing style:
rather than having~$\vpost$ in the post-condition of the chained triple, we instead quantify
over an evaluation context $\ectxs$ that represents an arbitrary continuation to run after $\expr$.
This continuation is assumed to satisfy a chained tripled in which $\vpost$ occurs in the preconditions.
We additionally quantify over several assertions $\pre_l$, $\pre_r$, $\post_l$, and $\post_r$ that
are used
% in the pre/post-conditions
to represent additional resources used by the continuation.

\section{A Separation Logic for Verifying One Interleaving}
\label{sec:seqlogic}
We now return to \angelic,
our program logic verifying that
\emph{one} interleaving of a \lang program
is safe and terminates.
We first present the assertions of \angelic~(\cref{sec:angelic_assertions})
and then present selected reasoning rules~(\cref{sec:angelic_rules}).
\subsection{Assertions of \angelic}
\label{sec:angelic_assertions}

Assertions of \angelic are Iris assertions of type $\iProp$, written $\pre$.
The fractional points-to assertion of \angelic takes the form $\loc \fpointsto{\qp} \vals$
(while we reuse the syntax of the points-to assertion
from \musketeer, the two assertions are different---recall that \angelic and \musketeer are totally disjoint).
% %
% \angelic guarantees termination because it prevents
% eliminating the so-called later modality.

A key aspect of \angelic is that this logic has two reasoning modes.
First, the \emph{running mode} takes the form
$\run\expr\post$, where $\expr$ is the expression being logically ``run'' and $\post$ is a postcondition,
% that is, a predicate over expression of the form $\lambda\val.\,\pre$.
The assertion $\run\expr\post$ is close to a weakest-precondition~(WP).
In fact, it enjoys all the rules of a standard \SL WP.
However, the running mode enjoys additional rules that allow one to dynamically ``select'' and verify just one
interleaving.
This selection is made possible thanks to a second mode,
that we call the \emph{scheduler mode}.
The scheduler mode involves two key assertions.
First, $\goal$ is an opaque assertion,
intuitively representing the proof obligation to verify the whole program.
Second, the assertion $\yielded{\tid}{\expr}$ asserts
the ownership of the task $\tid$,
and that this task yielded facing expression $\expr$.

The logic satisfies the following soundness theorem:
\begin{theorem}[Soundness of \angelic]
\label{thm:angelic}
If $\runex{\expr}{\_}{\iTrue}$ holds,
then there exists a value $\val$ and a
store $\store$ such that
$\steprtc{\expr}{\emptyset}{\val}{\store}$.
\end{theorem}

\subsection{Reasoning Rules of \angelic}
\label{sec:angelic_rules}
\begin{figure}\small\centering\morespacingaroundstar
\begin{mathpar}
% \inferrule[A-Assert]{}
% {\runex{(\eassert\vtrue)}{\val}{\val=\vtrue}}
%
\inferrule[A-Alloc]
{\pure{0 \leq \ofs}}
{\runex{(\ealloc{\ofs})}{\val}{\exists\loc.\,\loc \pointsto \blockrepeat{\vint}{\vunit}}}

\inferrule[A-Load]
{\pure{0 \leq \ofs < \length\vals} \\ \loc \fpointsto{\qp} \vals }
{\run{(\eload{\loc}{\ofs})}{\wal.\,\pure{\wal = \vals(\ofs)} \star \loc \fpointsto{\qp} \vals}}

\inferrule[A-Bind]
{\runex\expr\val{\run{(\efillctx\ectx\val)}\post}}
{\run{(\efillctx{\ectx}{\expr})}{\post}}

\inferrule[A-Store]
{\pure{0 \leq \ofs < \length\vals} \\ \loc \fpointsto \vals }
{\run{(\estore{\loc}{\ofs}{\val'})}{\wal.\,\pure{\wal = \vunit} \star \loc \pointsto {\blockupd\vals\ofs{\val'}}}}

\inferrule[A-Call]
{\run{(\subst{f}{\vfun{f}{x}{\expr}}{\subst{x}{\val}{\expr}})}{\post}}
{\run{\ecall{(\vfun{f}{x}{\expr})}{\val}}{\post}}

\inferrule[A-Conseq]
{ \run\expr{\post'} \\\\
(\,\forall\val.\;\post'\,\val \wand \post\,\val\,)
}
{\run\expr{\post}}
\end{mathpar}
\vspace{1.5em}
\begin{mathpar}
\inferrule[Yield]
{\forall\tid.\;\yielded{\tid}{\expr} \wand (\forall\val.\,\yielded{\tid}{\val} \wand \post\,\val \wand \goal) \wand \goal}
{\run{\expr}{\post}}

\inferrule[Resume]
{\yielded{\tid}{\expr} \\
\runex\expr\val{\yielded{\tid}{\val} \wand \goal}}
{\goal}

\inferrule[Fork]
{\forall \tid_1\,\tid_2.\;\yielded{\tid_1}{\expr_1} \wand
\yielded{\tid_2}{\expr_2} \wand
(\forall \val_1\,\val_2.\, \yielded{\tid_1}{\val_1} \wand
\yielded{\tid_2}{\val_2} \wand \post\,(\val_1,\val_2) \wand \goal) \wand  \goal
}
{\run{(\epar{\expr_1}{\expr_2})}{\post}}
\end{mathpar}
\captionlabel{Selected Reasoning Rules of \angelic (extends \cref{fig:key:angelic})}{fig:angelic}
\end{figure}

\Cref{fig:angelic} presents the key reasoning rules allowing
the user to select and verify an interleaving.
These inference rules are at the $\iProp$ level: their premises
are implicitly separated by $\star$, and the implication
between the premise and the conclusion is stated as the entailement $\vdash$.

The upper part of \Cref{fig:angelic} showcases that the run mode
of \angelic is, for its sequential part,
similar to a standard \SL.
% For example. \RULE{A-Assert} requires the user to show
% the Boolean of an assert is true before proceeding.
\RULE{A-Alloc} performs an allocation, \RULE{A-Load} a load
and \RULE{A-Store} a store---here, the allocation size and various offsets must be valid.
\RULE{A-Call} verifies a function call.
\RULE{A-Conseq} shows that the user can make the postcondition stronger.
% In a standard way, \RULE{A-Conseq} allows for proving a framing rule for \angelic.
%
\RULE{A-Bind} allows for reasoning under an evaluation context.

The lower part of \Cref{fig:angelic} focuses on the scheduler mode of \angelic.
\RULE{Yield} asserts (reading the rule from bottom to top)
that the proof of $\run{\expr}{\post}$ can pause,
and switch to the scheduler mode---that is, a proof where the target is $\goal$.
To prove this target,
the user gets to assume that some (universally quantified) task $\tid$
yielded with expression $\expr$, and that
when this expression will have reduced to a value $\val$ satisfying $\post$,
then $\goal$ will hold.

\RULE{Resume} is the companion rule of \RULE{Yield}: it asserts that in order
to prove $\goal$, the user has to give up the ownership of a task $\tid$
facing an expression $\expr$ and switch back to the running mode to verify
that $\runex\expr\val{\yielded{\tid}{\val} \wand \goal}$.

\RULE{Fork} shows the real benefit of the scheduler mode.
This rule asserts that, for verifying the parallel primitive
$\epar{\expr_1}{\expr_2}$, the user can switch to the scheduler mode.
In this mode, the user gets to suppose that two tasks
$\tid_1$ and $\tid_2$
yielded at $\expr_1$ and $\expr_2$, respectively.
Moreover, the user can suppose that, when these two tasks would have completed
their execution and reached values $\val_1$ and $\val_2$ such that
$\post\,(\val_1,\val_2)$ hold, the $\goal$ will hold.
At this point, the user can choose which of $\expr_1$ and $\expr_2$ to begin verifying using \RULE{Resume}.

Recall in \Cref{sec:key:angelic} we saw rules \RULE{A-ParSeqL} and \RULE{A-ParSeqR} allowing
one to verify a parallel composition by picking either a left-then-right or right-then-left sequential ordering.
These two rules can be derived from the more general constructs of \angelic that we have now seen.
For example, in order to show that \RULE{A-ParSeqL} holds,
we first apply \RULE{Fork}, then use \RULE{Resume} for the expression $\expr_1$.
We then use \RULE{A-Conseq} with \RULE{Resume} for expression $\expr_2$ and conclude.

\section{Case Studies}
\label{sec:case_studies}
To showcase \musketeer,
we start by using it to prove the soundness of a simple affine
type system that ensures schedule-independent safety~(\cref{sec:toy}).
We then extend this type system with two core algorithmic primitives proposed by \citet{bfgs12-pbbs} for ensuring internal determinacy: priority writes~(\cref{sec:prio}) and deterministic hash sets~(\cref{sec:hash}).
Interestingly, while these primitives appear to be internally deterministic to their client,
their implementation is not: they observe internal
state of shared data structures that may be concurrently modified.
Yet, we show that they satisfy schedule-independent safety.
Because all well-typed programs in this system have schedule-independent safety, we can use \angelic to reason about them, as we demonstrate by verifying a parallel list deduplication example~(\cref{sec:dedup}).

\subsection{\typesystem: An Affine Type System for Determinism}
\label{sec:toy}

This section presents \typesystem, an affine type system for \lang that ensures determinism.
Like many other substructural type systems, the types in \typesystem can be thought of as tracking
\emph{ownership} of resources such as array references, thereby preventing threads from accessing
shared resources in a way that would introduce nondeterministic behaviors.

\paragraph{Syntax}
\newcommand{\tempty}{\textsf{empty}}
\newcommand{\tinvalid}{\bot}
\newcommand{\vars}{\textsf{Var}}

\begin{figure}\small\centering
\[\begin{array}{r@{\;\;}l}
\tau &\eqdef\; \tinvalid \mid \tempty \mid \tunit \mid \tbool \mid \tint \mid \tarrow{\tau}{\tau} \mid \tprod{\tau}{\tau} \mid \tref\tau \\
\Gamma &\in\; \vars \rightharpoonup \tau\\
\end{array}\]
\[\begin{array}{r@{\;\;\eqdef\;\;}l@{\qquad}r@{\;\;\eqdef\;\;}l}
\tempty \cdot \tempty & \tempty & \tint \cdot \tint&\tint \\
\tunit \cdot \tunit&\tunit &\tprod{\tau_1}{\tau_2} \cdot \tprod{\tau_1'}{\tau_2'} & \tprod{(\tau_1 \cdot \tau_1')}{(\tau_2 \cdot \tau_2')}\\
\tbool \cdot \tbool&\tbool & (\tarrow{\tau_1}{\tau_2}) \cdot (\tarrow{\tau_1'}{\tau_2'}) & \eif{(\tau_1=\tau_1' \wedge \tau_2=\tau_2')}{\tarrow{\tau_1}{\tau_2}}{\tinvalid}
\end{array}\]
\captionlabel{Syntax of \minidet Type System}{fig:linsyntax}
\end{figure}

The syntax of types in \typesystem appears in \Cref{fig:linsyntax}.
A type $\tau$ is either the invalid type~$\tinvalid$ (used only internally),
the $\tempty$ type, describing an unknown value without ownership,
the unit type~$\tunit$, the Boolean type $\tbool$, the integer type $\tint$,
the arrow type $\tarrow{\tau_1}{\tau_2}$,
the immutable product~$\tprod{\tau_1}{\tau_2}$ or the reference type $\tref{\tau}$.
A typing environment $\Gamma$ is a finite map from variables to types.
We write $\dom\Gamma$ for its domain.

The type system is \emph{affine}
meaning that, when splitting a typing context in two,
a variable can only appear in one sub-context at a time.
However, variables
with types whose inhabitants have no associated notion of ownership,
or variables with types with fractional reasoning,
can be split and joined.
In order to capture this notion, we
equip types with a monoid operation
$\_ \cdot \_$ taking two types as arguments
and producing a new type.
In particular, when $\tau \cdot \tau = \tau$, it means
that a variable of type $\tau$ can be duplicated.
The definition of the monoid operation appears in the lower part of \Cref{fig:linsyntax}.
The missing cases are all sent to $\tinvalid$.
In particular these definitions prevent a reference from being duplicated.
We extend the monoid operation to typing environments by defining
$\Gamma_1 \cdot \Gamma_2$ as the function that maps the variable $x$ to
$\tau_1$ if $\Gamma_1(x)=\tau_1$ and $x$ is not in the domain of $\Gamma_2$,
$\tau_2$ if $\Gamma_2(x)=\tau_2$ and $x$ is not in the domain of $\Gamma_1$,
and $\tau_1 \cdot \tau_2$ if $\Gamma_1(x)=\tau_1$ and $\Gamma_2(x)=\tau_2$.

The typing judgement of \typesystem
takes the form
$\judg{\Gamma}{\expr}{\tau}{\Gamma'}$,
and asserts that $\expr$ has type $\tau$ and transforms the
typing environment $\Gamma$ into $\Gamma'$.

\paragraph{Typing rules}
\begin{figure}
\begin{mathpar}
\inferrule[T-Var]{}
{\judg{\msingleton{x}{\tau}}{x}{\tau}{\emptyset}}

\inferrule[T-Unit]{}
{\judg{\emptyset}{\vunit}{\tunit}{\emptyset}}

\inferrule[T-Bool]{}
{\judg{\emptyset}{\vbool}{\tbool}{\emptyset}}

\inferrule[T-Int]{}
{\judg{\emptyset}{\vint}{\tint}{\emptyset}}

\inferrule[T-Assert]
{\judg{\Gamma}\expr\tbool{\Gamma'}}
{\judg{\Gamma}{\eassert\expr}{\tunit}{\Gamma'}}

\inferrule[T-Let]
{\judg{\Gamma_1}{\expr_1}{\tau_1}{\Gamma_1'} \\\\
\judg{\minsert{x}{\tau_1}{\Gamma_1'}}{\expr_2}{\tau_2}{\Gamma_2}
}
{\judg{\Gamma_1}{\elet{x}{\expr_1}{\expr_2}}{\tau_2}{\mdelete{x}{\Gamma_2}}}

\inferrule[T-Weak]
{\Gamma_1 \subseteq \Gamma_1' \\ \Gamma_2 \subseteq \Gamma_2'\\\\ \judg{\Gamma_1'}{\expr}{\tau}{\Gamma_2'}}
{\judg{\Gamma_1}{\expr}{\tau}{\Gamma_2}}

\inferrule[T-Abs]
{\Gamma = \Gamma \cdot \Gamma \\\\ \judg{\minsert{f}{\tarrow{\tau}{\tau'}}{\minsert{x}{\tau}{\Gamma}}}{\expr}{\tau'}\emptyset}
{\judg\Gamma{\efun{f}{x}{\expr}}{\tarrow{\tau}{\tau'}}\emptyset}

\inferrule[T-App]
{\judg{\Gamma_1}{\expr_1}{\tau}{\Gamma_2}\\\\
\judg{\Gamma_2}{\expr_2}{\tarrow{\tau}{\tau'}}{\Gamma_3}}
{\judg{\Gamma_1}{\ecall{\expr_2}{\expr_1}}{\tau'}{\Gamma_3}}

\inferrule[T-Ref]
{\judg\Gamma\expr\tau{\Gamma'}}
{\judg\Gamma{\eref{\expr}}{\tref{\tau}}{\Gamma'}}

\inferrule[T-Get]{}
{\judg{\msingleton{x}{\tref\tau}}{\eget{x}}{\tau}{\msingleton{x}{\tref\tempty}}}

\inferrule[T-Set]
{\judg\Gamma\expr{\tau}{\msingleton{x}{\tref\tempty} \cdot \Gamma'}}
{\judg\Gamma{\eset{x}{\expr}}{\tunit}{\msingleton{x}{\tref\tau} \cdot \Gamma'}}

\inferrule[T-Par]
{\judg{\Gamma_1}{\expr_1}{\tau_1}{\Gamma_1'} \\
\judg{\Gamma_2}{\expr_2}{\tau_2}{\Gamma_2'}
}
{\judg{\Gamma_1 \cdot \Gamma_2}{\epar{\expr_1}{\expr_2}}{\tprod{\tau_1}{\tau_2}}{\Gamma_1' \cdot \Gamma_2'}}

\inferrule[T-Frame]
{\judg{\Gamma}{\expr}{\tau}{\Gamma'}}
{\judg{\Gamma_0 \cdot \Gamma}\expr\tau{\Gamma_0 \cdot \Gamma'}}
%
% \inferrule[T-PAlloc]
% {\judg{\Gamma}{\expr}{\tint}{\Gamma'}}
% {\judg{\Gamma}{\palloc{\expr}}{\tpwrite{1}}{\Gamma'}}
%
% \inferrule[T-PWrite]
% {\judg\Gamma\expr{\tint}{\msingleton{x}{\tpwrite{q}} \cdot \Gamma'}}
% {\judg\Gamma{\pwrite{x; \expr}}{\tunit}{\msingleton{x}{\tpwrite{q}} \cdot \Gamma'}}
    %
    % \inferrule[T-Pread]{}
    % {\judg{\msingleton{x}{\tpread{q}}}{\pread{x}}{\tint}{\msingleton{x}{\tpread{q}}}}
    %
\end{mathpar}
\captionlabel{Selected Typing Rules of \minidet}{fig:linear}
% \todo{Add pairs?}
\end{figure}

Selected typing rules appear in \Cref{fig:linear}.
\RULE{T-Var} types variable $x$ at type $\tau$ if $x$ has type $\tau$
in the typing environment.
The returned environment is empty.
\RULE{T-Unit}, \RULE{T-Bool} and rule \RULE{T-Int} type unboxed values.
\RULE{T-Assert} types an assert primitive.
\RULE{T-Let} types a let-binding $\elet{x}{\expr_1}{\expr_2}$
at type $\tau_2$
with initial context $\Gamma_1$ if
$\expr_1$ has type~$\tau_1$ under the same context and produces context $\Gamma_2$,
and if $\expr_2$ has type $\tau_2$ under the context $\Gamma_1$ in which $x$ has type $\tau_1$.
The produced context of the let-binding is $\Gamma_2$ from which $x$ has been deleted.
\RULE{T-Abs} types a function with recursive name~$f$, argument $x$ and body $\expr$,
of type $\tau \rightarrow \tau'$
and with typing environment $\Gamma$.
This environment must be duplicable, that is $\Gamma = \Gamma \cdot \Gamma$.
This duplicability implies that $\Gamma$ contains no
types with ownership, that is, for now, no references.
The precondition requires that $\expr$ has type $\tau'$ in $\Gamma$,
augmented with~$f$ of type $\tau \rightarrow \tau'$ and $x$ of type $\tau$.
\RULE{T-App} types a function call and is straightforward.
\RULE{T-Ref} types a reference allocation.
\RULE{T-Get} types a get operation on a variable $x$.
This rule requires that $x$ is of some type $\tref\tau$, returns a type $\tau$
and updates the binding of $x$ to $\tref\tempty$.
This is because get returns the ownership
of the content of the cell---meaning
that the cell does not hold recursive ownership of its contents anymore.%
%The type $\tempty$ witnesses an object without any associated ownership.%
\footnote{We could have derived another rule for get on a reference whose content is not tied to any ownership. We follow this approach when extending the type system with priority writes~(\cref{sec:prio}).}
\RULE{T-Set} is the dual, and types the expression $\eset{x}{\expr}$.
This rule requires that $\expr$ is of some type $\tau$ and
that, in the resulting environment, $x$ is of type $\tref\tempty$.
The set operation returns unit and updates
the type of $x$ to $\tref\tau$, ``filling''
the cell.
\RULE{T-Par} types a parallel primitive,
and is similar to the related \SL rules.
Indeed, \RULE{T-Par} requires splitting the context in two parts,
that will be used to type separately the two sub-tasks,
whose result typing context will be merged in the
result typing context of the rule.
Finally, \RULE{T-Frame} allows for framing a part of the context for local reasoning,
and \RULE{T-Weak}
allows for removing bindings from the input and output typing environments.

\paragraph{Soundness of \minidet}
The above system prevents data-races, and hence guarantees that well-typed programs have schedule-indepedent safety,
as formalized by the following lemma.

\begin{lemma}[Soundness of \minidet]\label[lemma]{thm:type}
If $\;\judg{\emptyset}{\expr}{\tau}{\emptyset}$ holds,
then $\sis{\expr}$ holds.
\end{lemma}

% \paragraph{Semantic soundness proof}
\newcommand{\shape}{s}
\newcommand{\sinvalid}{\textsf{sinvalid}}
\newcommand{\snone}{\textsf{snone}}
\newcommand{\sref}[2]{\textsf{sref}\,#1\,#2}
\newcommand{\sarrow}[1]{\textsf{sarrow}\,#1}
\newcommand{\sprod}[2]{\textsf{sprod}\,#1\,#2}
\newcommand{\senv}{M}
\newcommand{\svars}{V}

\newcommand{\interp}[3]{\llbracket\,#1\mid #2 \mid #3 \,\rrbracket}
\newcommand{\interpone}[1]{\llbracket\,#1\,\rrbracket}

\newcommand{\savedprop}[2]{#1 \Mapsto #2}
\newcommand{\onlyleft}[1]{\textsf{onlyleft}\,#1}
\newcommand{\msubst}[2]{[#1/]#2}
\newcommand{\similartyp}[2]{#1 \approx #2}
\newcommand{\similarshape}[2]{#1 \approx #2}
\newcommand\restrict[1]{\raisebox{-.5ex}{$|$}_{#1}}

\begin{figure}\centering\small\morespacingaroundstar
\[
\shape \;\eqdef\; \sinvalid \mid \snone \mid \sprod\shape\shape \mid \sref\val\shape \mid \sarrow\gamma
\qquad \senv \;\in\; \vars \rightharpoonup \tau
\qquad \svars \;\in\; \vars \rightharpoonup \Values
\]
\vspace{-0.5em}
\[\begin{array}{r@{\;\;\eqdef\;\;}l@{\qquad}r@{\;\;\eqdef\;\;}l}
\interp\tempty\snone\val&\iTrue & \interp\tbool\snone\val&\pure{\exists b.\,\val=b}\\
\interp\tunit\snone\val&\pure{\val=\vunit} & \interp\tint\snone\val&\pure{\exists i.\,\val=i}\\
\end{array}\]
\vspace{-0.5em}
\[\begin{array}{r@{\;\;}l}
\interp{\tprod{\tau_1}{\tau_2}}{\sprod{\shape_1}{\shape_2}}\val&\eqdef\;\exists \val_1\,\val_2.\;\pure{\val=\vprod{\val_1}{\val_2}} \star \interp{\tau_1}{\shape_1}{\val_1} \star \interp{\tau_2}{\shape_2}{\val_2}\\
\interp{\tref\tau}{\sref{\wal}\shape}\val&\eqdef\;\exists\loc.\;\pure{\val=\loc} \star \loc \pointsto [\wal] \star \interp{\tau}{\shape}\wal\\
\interp{\tarrow{\tau}{\tau'}}{\sarrow{\gamma}}\val&\eqdef\;\exists\vpre.\,\savedprop\gamma{\vpre} \star \always \vpre \;\star\\
&\phantom{\eqdef\;}\onlyleft{(\always\forall\wal\,\shape.\;\atriplemore{\later \vpre \star \interp{\tau}{\shape}\wal}{(\ecall{\val}{\wal})}{\val'}{\shape'}{\interp{\tau'}{\shape'}{\val'}})}\\
\interp{\Gamma}{\senv}{\svars} &\eqdef\; \pure{\dom \Gamma = \dom\senv = \dom\svars} \star \bigast{x \in \dom{\Gamma}}{\interp{\Gamma(x)}{\senv(x)}{\svars(x)}}\\[0.5em]
&\text{where\quad } \onlyleft({P}) \eqdef \lambda\vbool.\,\eif{\vbool}{(\vpre\ \vtrue)}{\iTrue}\\
\end{array}\]
\begin{align*}
&\interpone{\judg{\Gamma}{\expr}{\tau}{\Gamma'} } \;\eqdef\;\forall\senv\,\svars.\\
&\qquad\atriplemore{\interp\Gamma\senv\svars}{(\msubst\svars\expr)}{\val}{(\shape,\senv')}{\pure{\similartyp{\Gamma}{\Gamma'} \;\wedge\;\similarshape{\senv}{\senv'}} \star \interp{\tau}{\shape}{\val} \star \interp{\Gamma'}{\senv'}{\svars\restrict{\dom{\Gamma'}}}}
\end{align*}
\vspace{-0.5em}
\captionlabel{Semantic Interpretation of \minidet}{fig:interp}
\end{figure}

To prove this theorem, we use program logic-based \emph{semantic typing}~\citep{timany-et-al-24}.
With this technique, we associate a triple (in our case, a \musketeer triple)
to a typing judgement, and show that whenever the typing judgement holds, the corresponding triple is valid.
%The proof of the validity of each typing rule amounts to a proof in \SL,
The soundness theorem of the type system is then derived from the soundness of the underlying logic.

The \musketeer triple associated to a typing judgement
makes use of a \emph{logical relation}.
Typically, when using program logic-based semantic typing, a logical relation is a relation expressed in the assertions of the underlying logic that relates a type to a value it inhabits.
In our case, however, the logical relation involves three parameters:
a type, a value, and a \emph{shape}.
The shape captures the ``determinism'' of each type and will be used in connection with ghost return values.
For example, the shape of a reference
is the actual value stored in this reference,
and the shape of a function records that the function's environment is deterministic.

\Cref{fig:interp} defines the format of shapes.
A shape $\shape$ as either an invalid shape
(whose purpose is similar to the invalid type, as we equip shapes with a monoid operation),
the none shape, storing no information, the product shape $\sprod{\shape_1}{\shape_2}$,
the reference shape $\sref\val\shape$, where $\val$ represents the content of the reference
and $\shape$ the shape associated with $\val$ and finally the arrow shape $\sarrow{\gamma}$,
where $\gamma$ is the name of an Iris ghost cell~\citep{iris}.

The logical relation $\interp\tau\shape\val$, shown in \cref{fig:interp} then
relates a type $\tau$, a shape $\shape$, and a value~$\val$.
This relation is itself of type \vProp{}.
% Cases not in the figure are sent to the always false assertion.
%
Unboxed types are interpreted as expected, associated the with $\snone$ shape.
Products must be associated to the product shape and a product value,
and the interpretation must recursively hold.
For the reference type $\tref\tau$, the shape must be a reference shape $\sref{\wal}{\shape}$,
$\val$ must be a location $\loc$ such that~$\loc$ points to~$\wal$ and that recursively $\interp{\tau}{\shape}{\wal}$ holds.
Note that the interpretation of a reference expresses the ownership of the associated points-to.
Besides, the content of the reference $w$ is \emph{not} existentially quantified, but rather given by the shape.

The case of an arrow $\tarrow{\tau}{\tau'}$ is subtle and differs from the approach used in other program logic based logical relations.
In the usual approach, the interpretation of $\tarrow{\tau}{\tau'}$ says
that $\val$ is in the relation if for any $\wal$ in the interpretation of $\tau$, a Hoare triple of a certain form holds for the application $\ecall{\val}{\wal}$.
Unfortunately, this approach cannot be used directly with \musketeer.
The reason is that the usual approach exploits the fact that the underlying logic is higher-order and impredicative, so that a Hoare triple is itself an assertion that can appear in the pre/post-condition of another triple.
In contrast, in \musketeer, the assertions appearing in pre/post-conditions are $\vProp$s, but the triple itself is not a $\vProp$, it is an $\iProp$ in the underlying chained logic, as we saw in \cref{sec:chained}.

To work around this, we define an operation $\onlyleft$ that takes an $\iProp$ and coerces it into a $\vProp$ by requiring the proposition to only hold for the left-hand side.
Using this, the logical relation asserts that,
only in the left case, for any value $\wal$ and shape $\shape$,
a Hoare triple holds for $\ecall{\val}{\wal}$.
In this triple, the precondition requires $\interp{\tau}{\shape}{\wal}$, and the postcondition says that the result will satisfy the interpretation of $\tau'$.
The precondition additionally requires $\vpre$ to hold for some existentially quantified \vProp{} predicate $\vpre$.
( Technically, $\vpre$ is assumed to hold under a \emph{later modality} $\later$, but this detail can be ignored.)
% alexandre: I removed the footnote: it takes space than a plain sentence
This $\vpre$ will correspond to the resources associated with whatever variables from a typing environment the function closes over.
Thus, $\vpre$ is required to hold under the Iris \emph{persistent} modality $\always$, ensuring that the proposition is duplicable---recall that the typing rule \RULE{T-Abs} requires functions to close over only duplicable environments.
% What does this $\vpre$ correspond to?
% Since a function can close over an environment, and there
% This $\vpre$ will correspond to the facts that hold about the
Finally, there is one last trick: to ensure that this existential quantification over $\vpre$ can later be eliminated using \RULE{M-ElimExist},
the witness is made unique by using an Iris \emph{saved predicate} assertion, lifted at the \vProp{} level
and written $\savedprop\gamma\vpre$, which states that~$\gamma$ is the name of a ghost variable that stores the assertion $\vpre$.
The $\gamma$ here is bound as part of the shape $\sarrow \gamma$.
Since a ghost variable can only store one proposition, only one $\vpre$ can satisfy this assertion.

\Cref{fig:interp} then
defines the interpretation of a typing environment $\Gamma$, a shape environment
$\senv$ and a value environment~$\svars$, written $\interp{\Gamma}{\senv}{\svars}$
as the lifting per-variable $x$ of the logical relation.
Using this, we obtain the interpretation
of the typing judgement $\judg{\Gamma}{\expr}{\tau}{\Gamma'}$.
This interpretation universally quantifies
over a shape environment $\senv$ and a variable environment $\svars$,
and asserts a \musketeer triple.
The precondition is the interpretation of the environments,
and targets an expression $\msubst{\svars}{\expr}$, that is,
the expression $\expr$ with variables replaced by values as
specified by~$\svars$.
The postcondition binds a return value $\val$
as well as a ghost return value consisting
of a shape~$\shape$ and a shape environment $\senv'$.
The postcondition asserts that the two typing environment $\Gamma$ and~$\Gamma'$
are \emph{similar}, written $\Gamma \approx \Gamma'$ and that the shape environments
$\senv$ and $\senv'$ are also similar, with (overloaded) notation $\similarshape{\senv}{\senv'}$.
Intuitively these relations guarantee that
variables did not change in nature in environments (\eg. a reference stayed a reference, and a reference shape stayed a reference shape, even if the content may have changed).
We formally define these statements in~\citeappendix{appendix:similar}.
The postcondition finally asserts
that the return value is related to $\tau$ and $\shape$
and that the returned environment $\Gamma'$ is correct with $\senv'$ and the same variables $\svars$, dropping unneeded bindings.

With these definitions, we state the fundamental lemma of the
logical relation.
\begin{lemma}[Fundamental]\label[lemma]{thm:fundamental}
If $\;\judg{\Gamma}{\expr}{\tau}{\Gamma'}$ holds
then $\interpone{\judg{\Gamma}{\expr}{\tau}{\Gamma'}}$ holds too.
\end{lemma}

From this lemma, it is easy to prove the soundness of \minidet~(\Cref{thm:type}).
Let us suppose that $\judg{\emptyset}{\expr}{\tau}{\emptyset}$ holds.
We apply \Cref{thm:fundamental} and learn that $\interpone{\judg{\emptyset}{\expr}{\tau}{\emptyset}}$ holds too.
Unfolding definitions and applying \RULE{M-Conseq},
this fact implies that $\atriplemore{\iTrue}{\expr}{\_}{\_}{\iTrue}$ holds.
We conclude by applying the soundness of \musketeer~(\Cref{thm:musketeer}).

\subsection{Priority Writes}
\label{sec:prio}
In this section, we extend \minidet with rules for \emph{priority writes}~\citep{bfgs12-pbbs}.
A priority write targets a reference $r$ on an integer~$x$ and
atomically updates the content $y$ of $r$ to $x \max y$.
As long as there are no concurrent reads, priority
writes can happen in parallel: because $\max$
is associative and commutative,
the order in which the parallel write operations happen does not matter.
Conversely, so long as there are no ongoing concurrent writes, reads from the reference will be safe and deterministic---and such reads can also happen in parallel.
Thus, priority writes are deterministic so long as they are used in a \emph{phased} manner, alternating between concurrent writes in one phase, and concurrent reads in the next.
For simplicity, we consider priority writes on integers equipped with the $\max$ function.

\paragraph{Implementation of priority writes}
\begin{figure}\centering\small
\[\begin{array}{r@{\;\;\eqdef\;\;}l@{\qquad\qquad}r@{\;\;\eqdef\;\;}l}
\palloc &\efunnonrec{n}{\eref\,n} & \preadname & \efunnonrec{r}{\eget\,r}\\
\pwritename & \multicolumn{3}{l}{\kern-0.5em \efun{f}{r\,x}{\elet{y}{\eget{r}}{\eif{x < y}{\vunit}{\eif{\ecas{r}{0}{x}{y}}{\vunit}{f\,r\,x}}}}}
\end{array}\]
\captionlabel{Implementation of Priority Writes}{fig:prio_code}
\end{figure}

\Cref{fig:prio_code} shows the implementation of
priority references.
Allocating a priority reference with $\pallocname$
just allocates a reference.
The priority read $\preadname$ is just a plain get operation.
A priority write $\pwritename$
is a function with recursive name $f$ taking two arguments:
$r$, the reference to update, and $x$, the integer to update the reference with.
The function tests if the content $y$ of the reference is greater than $x$.
If $x < y$, the function returns, because $x \max y = y$.
Else, the function attempts to overwrite $y$ with $x$ in $r$
with a CAS, and loops if it fails.

As noted by \citet{bfgs12-pbbs}, if we break the abstractions of the priority reference,
the implementation of $\pwritename$
is not internally deterministic: because $\pwritename$ reads $r$,
a location that can be written by a parallel task,
different interleavings might see different values.
However, because $\pwritename$
is carefully designed, these nondeterministic observations
are not externally visible and do not impact the safety of the program.
As we will see,
this latter fact allow us to derive a \musketeer
triple API to priority writes.
However, because nondeterminism is involved internally in the implementation,
we conduct the proof at the level of \chainedlog.

\paragraph{Extension of \minidet}
\newcommand{\tyupd}[2]{#1 \leadsto #2}
\begin{figure}\centering\small
\[\begin{array}{r@{\;\;\eqdef\;\;}l@{\qquad}r@{\;\;\eqdef\;\;}l}
\tau &  \multicolumn{3}{l}{\kern-0.5em \cdots \mid \tpwrite{\qp} \mid \tpread{\qp}}\\
\tpwrite{\qp_1} \cdot \tpwrite{\qp_2} & \tpwrite{(\qp_1+\qp_2)} & \tpread{\qp_1} \cdot \tpread{\qp_2} & \tpread{(\qp_1+\qp_2)}
\end{array}\]
\begin{mathpar}
\inferrule*[Left=T-PAlloc]
{\judg{\Gamma}{\expr}{\tint}{\Gamma'}}
{\judg{\Gamma}{\palloc{\expr}}{\tpwrite{1}}{\Gamma'}}

\kern1em\inferrule*[Left=T-PWrite]
{\judg{\Gamma}{\expr}{\tint}{\Gamma'} \\ \Gamma'(x) = \tpwrite\qp}
{\judg{\Gamma}{\expr}{\pwrite{x}\expr}{\Gamma'}}

\inferrule[T-PRead]{}
{\judg{\msingleton{x}{\tpread{q}}}{\pread{x}}{\tint}{\msingleton{x}{\tpread{q}}}}

\inferrule[T-Update]
{\tyupd\Gamma{\Gamma'} \\ \judg{\Gamma'}{\expr}{\tau}{\Gamma''}}
{\judg{\Gamma}{\expr}{\tau}{\Gamma''}}

\inferrule[U-Refl]{}{\tyupd{\tau}{\tau}}

\inferrule*[Left=U-Pair]
{\tyupd{\tau_1}{\tau_1'} \\ \tyupd{\tau_2}{\tau_2'}}
{\tyupd{\tprod{\tau_1}{\tau_2}}{\tprod{\tau_1'}{\tau_2'}}}

\inferrule[U-R2W]{}
{\tyupd{\tpread{1}}{\tpwrite{1}}}

\inferrule[U-W2R]{}
{\tyupd{\tpwrite{1}}{\tpread{1}}}
\end{mathpar}
\captionlabel{Extension of \minidet with Priority Writes}{fig:prio}
\end{figure}

\Cref{fig:prio} shows how we extend our type system.
We add two new type constructors,
$\tpwrite{\qp}$ and $\tpread{\qp}$,
asserting that the reference is in a write phase
with fraction $\qp$ or a read phase with fraction $\qp$, respectively.
The monoid on types is extended to sum fractions.
This definition implies, as we will see, that writes can happen in parallel with writes, and reads can happen in parallel with reads.

The lower part of \Cref{fig:prio} shows the new typing rules.
\RULE{T-PAlloc} allocates a priority reference and returns a type $\tpwrite{1}$.
\RULE{T-PWrite} types a priority write on some reference $x$ bound to the type $\tpwrite{\qp}$.
In particular, this rule does \emph{not} require the full fraction $1$,
meaning that the write operation can happen in parallel of other write operations.
\RULE{T-PRead} types a read similarly.
Again this rule does not require the full fraction.
\RULE{T-Update} allows for updating a typing context $\Gamma$ into $\Gamma'$
as long as $\tyupd{\Gamma}{\Gamma'}$.
This relation is defined pointwise over the elements
of the environments as the update relation
$\tyupd{\tau}{\tau'}$ which is defined last in \Cref{fig:prio}.
\RULE{U-Refl} asserts that a type can stay the same,
\RULE{U-Pair} distributes over pairs,
\RULE{U-R2W} transforms a read type into a write one, if the fraction is the full permission 1.
This precondition on the fraction is important:
it asserts that no parallel task use the priority reference.
\RULE{U-W2R} is symmetrical.

\paragraph{Extending the soundness proof}
\newcommand{\iswritename}{\textsf{ispw}\xspace}
\newcommand{\iswrite}[3]{\iswritename\,#1\,#2\,#3}
\newcommand{\isreadname}{\textsf{ispr}\xspace}
\newcommand{\isread}[3]{\isreadname\,#1\,#2\,#3}

\newcommand{\encaps}[1]{\{#1\}}
\newcommand{\encapsp}[1]{\{\lambda r \,\_.\; #1\}}

\newcommand{\spwrite}[1]{\textsf{spwrite}\,#1}
\newcommand{\spread}[1]{\textsf{spread}\,#1}

\begin{figure}\centering\small
\[\begin{array}{@{}r@{\;}c@{\;}l}
\encaps{\iTrue} & \palloc{\ofs} & \encapsp{\iswrite\loc{1}\ofs}\\
\encaps{\iswrite\loc\qp{i}} & \pwrite\loc{j} & \encaps{\lambda r \,\loc.\;\pure{\val=\loc} \star \iswrite{\loc}\qp{(i \max j)}}\\
\encaps{\isread\loc\qp{i}} & \pread\loc & \encapsp{\pure{\val=i} \star \isread\loc{1}{i}}
\end{array}\]
\begin{minipage}{0.5\textwidth}
\[\begin{array}{r@{\;}c@{\;}l}
\iswrite\loc{(\qp_1+\qp_2)}{(i \max j)} &\dashv\vdash& \iswrite\loc{\qp_1}{i} \star \iswrite\loc{\qp_2}{j} \\
\isread\loc{(\qp_1+\qp_2)}{i} &\dashv\vdash& \isread\loc{\qp_1}{i} \star \isread\loc{\qp_2}{i}\\
\iswrite{\loc}{1}{i} &\dashv\vdash& \isread{\loc}{1}{i}\end{array}\]
\end{minipage}
\begin{minipage}{0.49\textwidth}
\[\begin{array}{r@{\;}c@{\;}l}
\multicolumn{3}{c}{\shape \;\eqdef\; \cdots \mid \spwrite\vint \mid \spread\vint}\\
\interp{\tpwrite{\qp}}{\spwrite{i}}{\val} &\eqdef& \exists \loc.\, \pure{\val=\loc} \star \iswrite{\loc}{\qp}{i}\\
\interp{\tpread{\qp}}{\spwrite{i}}{\val} &\eqdef& \exists \loc.\, \pure{\val=\loc} \star \isread{\loc}{\qp}{i}
\end{array}\]
\end{minipage}
\captionlabel{Specifications of Priority Writes and Logical Interpretation}{fig:prio_proof}
\end{figure}

To extend the soundness proof to support these new rules, we first
prove specifications for the priority reference operations in \musketeer, shown in the upper part of \Cref{fig:prio_proof}.
These specifications involve
two predicates: $\iswrite{\loc}{\qp}{i}$,
asserting that $\loc$
is a priority reference, and
that $\loc$ is in its concurrent phase with fraction $\qp$
and stores (at least) $i$.
Symmetrically, $\isread\loc\qp{i}$ asserts that $\loc$ is in its read phase.
The specification of $\palloc{i}$ asserts that this function call returns a location $\loc$
such that $\iswrite{\loc}{\qp}{1}$ holds.
The specification of $\pwrite{\loc}{j}$ updates a share $\iswrite{\loc}{\qp}{i}$ into $\iswrite{\loc}{\qp}{(i \max j)}$.
The specification of $\pread{\loc}$ asserts that this function call returns the content of a
priority reference, if this reference is in its read phase.

The central part of \Cref{fig:prio_proof} shows the splitting
and joining rules of the \iswritename and \isreadname assertions.
It also shows that one can update a \iswritename assertion
into a \isreadname assertion, and vice-versa, as long as the fraction in 1
(formally, these conversions involve the so-called \emph{ghost updates}~\citep{iris}).

The lower part of \Cref{fig:prio_proof} intuits how we extend the logical
relation backing the soundness of our type system.
We add two shapes, one for each phase.
We then extend the logical relation as expected, making use of the previous assertions.

\subsection{Deterministic Concurrent Hash Sets}
\label{sec:hash}
Next, we extend \minidet with
a \emph{deterministic concurrent hash set}, inspired
by \citet{DBLP:conf/spaa/ShunB14}.
This hash set allows for concurrent, lock-free insertion,
and offers a function $\helemsname$ that returns an array with
the inserted elements in some arbitrary but deterministic order.
This hash set is implemented as an array,
and makes use of \emph{open addressing} and \emph{linear probing}
to handle collision.
The key idea to ensure determinism
is that neighboring elements in the array are \emph{ordered}
according to a certain total order relation.
As we will see, insertion preserves the ordering,
which in turn ensures determinism of the contents of the array.
\citet{DBLP:conf/spaa/ShunB14} also propose a deletion function,
which we do not verify.
The hash set usage must be \emph{phased}: insertion is allowed
to take place in parallel as long as no task
calls the function $\helemsname$.

\paragraph{Implementation of our hash set}
\begin{figure}\centering\small
\begin{minipage}{0.33\textwidth}
\begin{align*}
&\allocfillname \;\;\eqdef\;\;\lambda n\,v.\;\fillname\,(\ealloc{n})\,v\\[0.5em]
&\hinitname \;\;\eqdef\;\; \lambda h\,n.\\
&\quad \eassert{(n \geq 0)};\\
&\quad \eletprefix{d}{\eref{\vunit}}\\
&\quad \eletprefix{a}{\allocfill{n}{d}}\\
&\quad (a,d,h)\\[0.5em]
&\helemsname \;\;\eqdef\;\; \lambda (a,d,h).\\
&\quad\filtercompactname\,a\,d
\end{align*}
\end{minipage}
\begin{minipage}{0.65\textwidth}
\begin{align*}
&\haddname \;\;\eqdef\;\; \lambda (a,d,h)\,x.\\
&\quad \textsf{let}\,put\,=\,\mu f\,x\,i.\\
&\qquad \eletprefix{y}{\eload{a}{i}}\\
&\qquad \textsf{if}\,x==y\,\textsf{then}\,\vunit\,\textsf{else}\\
&\qquad \textsf{if}\,x==d \,\textsf{then}\,(\eif{\ecas{a}{i}{d}{x}}{\vunit}{f\,x\,i})\;\textsf{else}\\
&\qquad \eletprefix{j}{(i + 1) \,\mathsf{mod}\, (\elength{a})}\\
&\qquad \textsf{if}\,x < y\,\textsf{then}\,f\,x\,j\,\textsf{else} \,(\eif{\ecas{a}{i}{y}{x}}{f\,y\,j}{f\,x\,i})\, \textsf{in}\\
&\quad put\,x\,((h\,i)\, \mathsf{mod}\, (\elength{a}))
\end{align*}
\end{minipage}
\captionlabel{Implementation of a Deterministic Concurrent Hash Set}{fig:hashset}
\end{figure}

\Cref{fig:hashset} presents the implementation of the deterministic
hash set.
While in our mechanization we support a hash set over arbitrary values,
for space constraints we present here an implementation specialized to integers,
equipped with the comparison function~$<$.

A new hash set is initialized with the function $\hinit{h}{n}$,
which returns a tuple $(a,d,h)$,
where $a$ is the underlying array, $d$ is a dummy element
(in our case, a fresh reference containing the unit value) representing
an empty slot in the array.
The function $h$ is the hash function.
The implementation uses a helper routine, $\allocfill{n}{d}$, which allocates an array and fills it
with the value $v$ using a function $\fillname$, which we omit for
brevity.
The function $\helems{(a,d,h)}$ returns
a fresh array containing the elements of $a$ obtained by filtering
those equal to the dummy element $d$.
The key challenge in the design is to ensure that this operation
will be deterministic: in conventional linear probing hash tables,
the order of elements in the array would depend on the order of
insertions, so concurrent insertions would lead to nondeterministic orders.

To avoid this nondeterminism, the function
$\hadd{(a,d,h)}{x}$, which inserts $x$ in the hash set $(a,d,h)$,
enforces an ordering on elements in the array according
to the comparison function $<$.
The code makes use of a recursive auxiliary function $put$,
parameterized by an element $x$ and an index $i$, which tries to insert $x$ at $i$.
The function $put$ loads the content of the array $a$ at offset $i$
and names it~$y$. If $y$ is equal to $x$, then $x$
is already in the set and the function returns.
If $y$ is equal to the dummy element, the function tries a CAS
to replace $y$ with~$x$, and loops in case the CAS fails.
Otherwise, $y$ is an element distinct from $x$.
The function names the next index $j = (i + 1) \,\mathsf{mod}\,(\elength{a})$
and tests if $x < y$.
If $y$ is greater than $x$,
the function tries to insert $x$ at the next index $j$ by doing a recursive call of $f\,x\,j$.
If $x$ is greater than $y$,
the function tries to replace
$y$ with $x$ with a CAS, and loops if the CAS fails.
If the CAS succeeds, the function removed $y$ from the hash set,
and must hence insert it again by doing a recursive call $f\,y\,j$.

The function $\haddname$ then simply calls $put$ to insert $x$ at the initial
index $(h\,x) \,\textsf{mod}\, (\elength{a})$.

\paragraph{Extension of \minidet}
\newcommand{\tintarrayname}{\textsf{intarray}\xspace}
\newcommand{\tintarray}[1]{\tintarrayname\,#1}
\newcommand{\tintsetname}{\textsf{intset}\xspace}
\newcommand{\tintset}[1]{\tintsetname\,#1}
\newcommand{\hspec}[1]{\forall x.\;\atriple{\iTrue}{#1\,x}\val{\pure{\val=hash(x)}}}
\newcommand{\cspec}[1]{\forall x\,y.\;\sltriple{\iTrue}{c\,x\,y}{\lambda\val.\,\pure{\val=cmp(x,y)}}}
\newcommand{\totalorder}[1]{\textsf{TotalOrder}\,#1}
\begin{figure}
\[\begin{array}{r@{\;\;\eqdef\;\;}l}
\tau & \cdots \mid  \tintarray\qp \mid \tintset\qp\\
\tintarray{\qp_1} \cdot \tintarray{\qp_2} & \tintarray{(\qp_1+\qp_2)} \qquad\quad \tintset{\qp_1} \cdot \tintset{\qp_2} \;\eqdef\; \tintset{(\qp_1+\qp_2)}
\end{array}\]
\begin{mathpar}
\inferrule[T-AAlloc]
{\judg{\Gamma_1}{\expr_1}{\tint}{\Gamma_2} \\
\judg{\Gamma_2}{\expr_2}{\tint}{\Gamma_3}}
{\judg{\Gamma}{\allocfill{\expr_2}{\expr_1}}{\tintarray{1}}{\Gamma_3}}

\inferrule[T-ALoad]
{\judg{\Gamma_1}{\expr}{\tint}{\Gamma_2} \\ \Gamma_2(x) = \tintarray{\qp}}
{\judg{\Gamma_1}{\eload{x}{\expr}}{\tint}{\Gamma_2}}

\inferrule[T-AStore]
{\judg{\Gamma_1}{\expr_1}{\tint}{\Gamma_2} \\\\
\judg{\Gamma_2}{\expr_2}{\tint}{\Gamma_3} \\ \Gamma_3(x)=\tintarray{1}}
{\judg{\Gamma}{\estore{x}{\expr_2}{\expr_1}}{\tunit}{\Gamma_3}}

\inferrule[T-SAlloc]
{\judg{\Gamma}{\expr}{\tint}{\Gamma'} \\\\
% \totalorder{cmp} \\\\
(\hspec{h})
%(\cspec{c})
}
{\judg{\Gamma}{\hinit{h}{\expr}}{\tintset{1}}{\Gamma'}}

\inferrule*[Left=T-SAdd]
{\judg{\Gamma}{\expr}{\tint}{\Gamma'} \\
\Gamma'(x) = \tintset\qp
}
{\judg{\Gamma}{\hadd{x}{\expr}}{\tunit}{\Gamma'}}

\kern1em\inferrule*[Left=T-SElems]
{\judg{\Gamma}{\expr}{\tintset{1}}{\Gamma'}}
{\judg{\Gamma}{\helems{\expr}}{\tintarray{1}}{\Gamma'}}
\kern-1em
\end{mathpar}
\captionlabel{Extension of \minidet with Integer Arrays and Hash Set}{fig:hashset_typ}
\end{figure}

\Cref{fig:hashset_typ} presents the
extension of \minidet with this hash set.
To avoid issues related to ownership of the elements in the set,
% for simplicty,
we consider a hash set containing integers.

We add two new types: $\tintarray{\qp}$ describing an
array of integers with a fraction $\qp$ and $\tintset{\qp}$
a hash set of integers with a fraction $\qp$.
The monoid on types is extended to sum the fractions.

\RULE{T-AAlloc} types the allocation of an array filled with a default element.
% that is, a call to $\allocfillname$.
\RULE{T-ALoad} types a load operation on an array bound to the variable $x$.
This operation requires any fraction of $\tintarrayname$.
\RULE{T-AStore} types a store operation but
requires full ownership of the array---that is, the fraction 1.

\RULE{T-SAlloc} allocates a hash set.
This rule has one non-syntactical precondition,
which cannot be handled by a type system.
It requires that the hash function $h$, the first parameter of $\haddname$,
implements some arbitrary pure function $hash : \Values \to Z$.
This proof can be derived in \musketeer, and ensures that calls to the hash function are deterministic.
\RULE{T-SAlloc} returns a $\tintsetname$ type with fraction 1.

\RULE{T-SAdd} types an $\haddname$ operation on
a hash set $x$ with an arbitrary fraction $q$, meaning
that this operation can happen in parallel.
\RULE{T-SElems} types the $\helemsname$ operation,
requiring the full ownership of a hash set,
and producing a fresh array.
This operation consumes the hash set argument;
this is for simplicity: the hash set is only read and
is in fact preserved by the operation.

\paragraph{Extending the soundness proof}
\newcommand{\ishashsetname}{\textsf{hashset}\xspace}
\newcommand{\ishashset}[3]{\ishashsetname\,#1\,#2\,#3}
\newcommand{\caninsert}[2]{\textsf{caninsert}\,#1\,#2}
\newcommand{\singleton}[1]{\{#1\}}
\newcommand{\sintset}[1]{\textsf{sintset}\,#1}

\begin{figure}\centering\small\morespacingaroundstar
\begin{minipage}{0.3\textwidth}
\begin{mathpar}
\inferrule
{(\hspec{h})
%  \\
% (\cspec{c}) \\
% \totalorder{cmp}
}
{\atriple{\iTrue}{\hinit{h}{\ofs}}{\val}{\ishashset\val{1}{\emptyset}}}
%
% \inferrule
% {%\caninsert\val{x}
% }
% {\atriple{\ishashset\val\qp{X}}{\hadd{\val}{x}}{\wal}{\pure{\wal=\vunit} \star \ishashset\val\qp{(\singleton{x} \cup X)}}}
%
% \inferrule{}
% {\atriplemore{\ishashset\val{1}{X}}{\helems{\val}}{\wal}{(\loc,\wals)}{\pure{\deduped{X}{\wals}} \star \loc \pointsto \wals}}
\end{mathpar}
\end{minipage}
\begin{minipage}{0.65\textwidth}
\[\begin{array}{@{}r@{\;}c@{\;}l}
\encaps{\ishashset\val\qp{X}} & \hadd{\val}{\ofs} & \encapsp{\ishashset\val\qp{(\singleton{\ofs} \cup X)}}\\
\encaps{\ishashset\val{1}{X}} & \helems{\val} & \encaps{\lambda \val' \,(\loc,\wals).\;\pure{\val'=\loc} \star \loc \pointsto \wals}
\end{array}\]
\end{minipage}
\[\ishashset{\val}{(\qp_1+\qp_2)}{(X_1 \cup X_2)} \;\dashv\vdash\;  \ishashset{\val}{\qp_1}{X_1} \star \ishashset{\val}{\qp_2}{X_2}\]
\[\shape \;\;\eqdef\;\; \cdots \mid \sintset{X} \qquad  \interp{\tintset{\qp}}{\sintset{X}}{\val} \;\;\eqdef\;\; \ishashset{\val}{\qp}{X} \]
\captionlabel{Specifications of a Deterministic Hash Set and Logical Interpretation}{fig:hashset_proof}
\end{figure}

The upper part of \Cref{fig:hashset_proof} presents the \musketeer
specifications of the hash set operations.
These specifications make use of an assertion $\ishashset{\val}{\qp}{X}$
asserting that $\val$ is a hash set with fraction $\qp$ and content at least $X$, a set of values.
When $\qp=1$, then~$X$ is exactly the set of values in the set.
The specification of $\hinit{h}{i}$ returns a fresh set
with fraction 1 and no elements, provided that the parameter $h$ behaves correctly.
The specification of $\hadd{v}{\ofs}$ verifies the insertion of an integer $\ofs$
in a hash set $v$ with an arbitrary fraction $\qp$ and current content $X$,
which the function call updates to $(\singleton{\ofs} \cup X)$.
Since we specialize to hash sets of integers, we know that the inserted value
will not be the dummy element.
In our mechanization, we offer a more general specification,
allowing the user to insert other pointers %---as long as they are able to exhibit the associated points-to assertion,
as long as they ensure that the inserted pointer is not the dummy element.
Perhaps most importantly, the specification of $\helems{v}$
consumes an assertion $\ishashset{\val}{1}{X}$
with fraction 1 and produces an array~$\loc$ with a \emph{deterministic}
content $\wals$.
% containing the elements of $X$ without duplicates.
%
\Cref{fig:hashset_proof} then gives the reasoning rule
for splitting a $\ishashsetname$ assertion, enabling
parallel use.

The lower part of \Cref{fig:hashset_proof} shows how we extend the
logical relation. We add a shape $\sintset{X}$, where
$X$ a set of integers.
The interpretation of $\tintset{\qp}$
with shape $\sintset{X}$ and value $\val$
is then simply $\ishashset{\val}{\qp}{X}$.

\subsection{Deduplication via Concurrent Hashing}
\label{sec:dedup}
\newcommand{\parforname}{\textsf{parfor}\xspace}
\newcommand{\parfor}[3]{\parforname\,#1\,#2\,#3}
\newcommand{\selfname}{f}
\newcommand{\lowbound}{i}
\newcommand{\highbound}{j}
\newcommand{\parforarg}{k}
\newcommand{\diffname}{(\highbound-\lowbound)}
\newcommand{\midname}{mid}
\newcommand{\dedupname}{\textsf{dedup}\xspace}
\newcommand{\fractional}[2]{\textsf{Fractional}\,#1\,#2}

\newcommand{\parforspecname}{\textsf{forspec}\xspace}
\newcommand{\parforspec}[4]{\parforspecname\,#1\,#2\,#3\,#4}

\newcommand{\ahashset}[3]{\textsf{ahashset}\,#1\,#2\,#3}

\begin{figure}\centering\small
\begin{minipage}{0.35\textwidth}
\begin{align*}
&\parforname\;\eqdef\;\hat\mu\selfname.\,\lambda \lowbound\,\highbound\,\parforarg.\\
&\quad\textsf{if}\; \eeq\diffname{0}\;\textsf{then}\;\vunit\\
&\quad\textsf{else\;if}\;\diffname == {1}\;\textsf{then}\;\parforarg\,\lowbound\\
&\quad\textsf{else}\;\textsf{let}\;\midname\;=\; \lowbound + (\diffname/2)\;\textsf{in}\\
&\qquad\epar{(\selfname\,\lowbound\,\midname\,\parforarg)}{(\selfname\, \midname\,\highbound\,\parforarg)}
\end{align*}
\end{minipage}
\begin{minipage}{0.37\textwidth}
% Definition dedup : expr :=
%   λ: "l",
%     let: "start" := 0 in
%     let: "length" := Length "l" in
%     let: "table" := htbl_alloc h c ("length" '+ 1) in
%     parfor "start" "length"  (λ: "i", let: "x" := "l".["i"] in htbl_insert "table" "x") ;;
%     let: "r" := htbl_elems "table" in
%     Pair "l" "r".
\begin{align*}
&\dedupname\;\eqdef\;\lambda h\,a.\\
&\quad\eletprefix{start}{0}\\
&\quad\eletprefix{len}{\elength{a}}\\
&\quad\eletprefix{s}{\hinit{h}{(len + 1)}}\\
&\quad\parfor{start}{len}{(\lambda\ofs.\,\hadd{s}{(\eload{a}{i})})};\\
&\quad\eprod{a}{(\helems{s})}
\end{align*}
\end{minipage}
\captionlabel{Implementation of parfor and dedup Functions}{fig:dedup}
\end{figure}

For our last example, we consider \emph{array deduplication}, one of the parallel benchmark problems proposed by \citet{bfgs12-pbbs}.
The task is to take an array of elements and return an array containing the same elements but with duplicates removed.
The solution proposed by \citet{bfgs12-pbbs} is to simply insert all the elements in parallel into a deterministic hash set and then return the elements of the hash set.
\Cref{fig:dedup} presents $\dedupname$, an implementation of this algorithm in \lang.
To do the parallel inserts, it uses a helper routine called $\parfor{i}{j}{k}$,
which runs $(k\, n)$ in parallel for all $n$ between $i$ and $j$.
Our goal is to prove that $\dedupname$ satisfies schedule-independent safety, and then prove a specification in \angelic.
Throughout this proof, we assume that we have some hash function $h$
such that $\forall x.\;\atriple{\iTrue}{(h\,x)}{\val}{\pure{\val=hash\,x}}$
and
$\forall x.\;\runex{(h\,x)}{\val}{\pure{\val=hash\,x}}$,
where $hash$ is some function in the meta-logic.

Our first step is to show that $\dedupname$ can be typed in \minidet.
This follows by using a typing rule for parfor (given in \citeappendix{appendix:hash:parfor}), and the earlier typing rules we derived for the hash set. Using these, we derive
$\judg{\emptyset}{\dedupname\,h}{\tarrow{\tintarray{\qp}}{\tprod{\tintarray{\qp}}{\tintarray{1}}}}{\emptyset}$.
Thus, for a well-typed input array $a$, $\dedupname\,h\,a$ satisfies schedule-independent safety.

We then verify $\dedupname$ using \angelic.
The proof uses \angelic reasoning rules for the hash set, shown in \citeappendix{appendix:hash:angelic}, which are
similar to the earlier \musketeer specifications~(\cref{sec:hash}), except for three key points.
First, the \angelic specification shows that, for a set $v$ with content $X$,
$\helems{v}$ returns an array $\wals$ which contains just the elements of the set $X$.
Second, the representation predicate for the hash set has no fraction:
there is never a need for splitting it in \angelic.
Third, as we require the user to prove termination, the representation
predicate tracks how many elements have been inserted, and does not allow inserting into a full table.

Finally, we use a derived specification for $\parfor{i}{j}{k}$ that allows us to reason about it as if it were a sequential for-loop:
\[\parforspec{i}{j}{k}{\pre} \;\vdash\; \run{(\parfor{i}{j}{k})}{\lambda\val.\,\pure{\val=\vunit} \star \pre}\]
Here, $\parforspec{i}{j}{k}\pre$ is defined recursively as
\[\morespacingaroundstar\parforspec{i}{j}{k}{\pre} \;\eqdef\; \big(\,\pure{i \geq j} \star \pre\,\big)  \;\lor\; \big(\,\pure{i < j} \star \runex{(k\,i)}{\val}{\pure{\val=\vunit} \star \parforspec{(i+1)}{j}{k}{\pre}}\,\big)\]
In this definition, either $i \geq j$ and the postcondition holds (since there are no recursive calls to be done),
or $i < j$, and the user has to verify $k\,i$,
and show that $\parforspec{(i+1)}{j}{k}{\pre}$ holds afterward.
Essentially, this generalizes the idea we saw earlier in \RULE{A-ParSeqL},
by having us verify an interleaving that executes each task
sequentially from $i$ to $j$.
With these specifications, we deduce the following \angelic specification
for $\dedupname$:
\[\morespacingaroundstar
\loc \fpointsto\qp \vals \;\vdash\;
\runex{(\dedupname\,h\,\loc)}{\val}
{\exists\loc'\,\wals.\; \loc \fpointsto\qp \vals \star \loc' \mapsto \wals \star \pure{\deduped{\wals}{\vals}}}
\]

\section{Related Work}
\label{sec:related_work}
\paragraph{Deterministic parallel languages}
As shown in \Cref{sec:toy}, \musketeer
can be used to prove the soundness of language-based techniques for enforcing determinism.
A large body of such techniques exist, and it would be interesting to apply \musketeer to some of these.
In general, these languages typically ensure determinism by
restricting side effects (e.g., in purely functional languages)
or by providing the programmer with fine-grained control over scheduling
of effects (e.g., in the form of a powerful type-and-effect system).
Examples include seminal works such as
Id~\cite{DBLP:journals/toplas/ArvindNP89} and
NESL~\cite{DBLP:journals/jpdc/BlellochHSZC94}
as well as related work on
Deterministic Parallel Java~\cite{DBLP:conf/oopsla/BocchinoADAHKOSSV09,DBLP:conf/popl/BocchinoHHAAWS11},
parallelism in
Haskell~\cite{DBLP:conf/fsttcs/JonesLKC08,DBLP:conf/icfp/KellerCLJL10,DBLP:conf/popl/ChakravartyKLMG11,DBLP:conf/europar/ChakravartyKLP01,DBLP:conf/haskell/MarlowNJ11},
the LVars/LVish framework~\cite{DBLP:conf/pldi/KuperTTN14,DBLP:conf/popl/KuperTKN14,DBLP:conf/icfp/KuperN13},
Liquid Effects~\cite{DBLP:conf/pldi/KawaguchiRBJ12}.
Manticore~\cite{DBLP:conf/popl/FluetRRSX07},
SAC~\cite{DBLP:journals/jfp/Scholz03},
Halide~\cite{DBLP:conf/pldi/Ragan-KelleyBAPDA13},
Futhark~\cite{DBLP:conf/pldi/HenriksenSEHO17},
and many others.

It is typically challenging to formally prove sequentialization or determinization results for these kinds of languages, particularly in an expressive language with features like higher-order state and recursive types.
For example, \citet{DBLP:conf/popl/Krogh-Jespersen17} point out that it took 25 years for the first results proving that in a type-and-effect system, appropriate types can ensure that a parallel pair is contextually equivalent to a sequential pair.
They show how a program-logic based logical relation, like the one we used in \Cref{sec:case_studies}, can vastly simplify such proofs.
\musketeer provides a program logic that is well-suited for constructing models to prove whole-language determinism properties.
Although not discussed in this paper, we have already completed
a proof of schedule-independent safety for a simplified model of the LVars
framework.
We believe similar results may be possible for other
deterministic-by-construction languages.

\paragraph{Logic for hyperproperties}
So-called \emph{relational}
program logics have been developed to prove hyperproperties.
\citet{DBLP:conf/isola/Naumann20} provides an extensive survey of these logics.
A number of such logics support very general classes of hyperproperties~\citep{DBLP:journals/pacmpl/DOsualdoFD22, DBLP:conf/pldi/SousaD16}.
However, most of the relational logics building on concurrent separation logic have been
restricted $\forall\exists$ hyperproperties~\citep{liang-feng-16,reloc-reloaded-21,simuliris-22,trillium}.
Because schedule-independent safety is a $\forall\forall$ property, it falls outside the scope of these logics, which motivated our development of \chainedlog.
In the world of $\forall\forall$ hyperproperties, several
Iris-based relational logics have been proposed.
For example, \citet{reloc-reloaded-21} and \citet{gregersen-et-al-21}
verify variations of \emph{non-interference} in a sequential setting.
Both works require that both executions terminate.
However, their underlying relational logics do not support
\musketeer's distinctive feature: the chaining rule \RULE{C-Chain}.
In another setting, \citet{timany-et-al-17} present a logical relation
showing that Haskell's ST monad~\citep{launchbury-peyton-jones-95} properly
encapsulates state.
They show such a result using a \emph{state-independence}
property which intuitively asserts
that, for a well-typed program,
if one execution terminates with a particular initial heap,
then every execution terminates with any other initial heap.

Most logics for hyperproperties are structured as relational logics.
However, some, like \musketeer, prove a hyperproperty through unary reasoning.
For example, \citet{DBLP:journals/pacmpl/DardinierM24},
target arbitrary hyperproperties
for a pure language, with a triple referring to a single expression, but
with pre/post-conditions describing multiple executions.
\citet{commcsl} present CommCSL,
a concurrent \SL for proving \emph{abstract commutativity},
that is, where two operations commute \emph{up-to} some abstract interface.
This idea appears for example in the API for priority writes,
which implies that writes commutes~(\cref{sec:prio}).
In contrast with our approach, CommCSL is globally parameterized by
a set of specifications the logic ensures commute.
In \musketeer, no such parameterization is needed:
proof obligations are entirely internalized.

\paragraph{Commutativity-Based Reasoning}
Schedule-independent safety reduces the problem of verifying safety for all executions of a program to just verifying safety of any one terminating execution.
This can be seen as an extreme form of a common technique in concurrent program verification, in which the set of possible executions of a program is partitioned into equivalence classes, and then a representative element of each equivalence class is verified~\citep{DBLP:conf/lics/Farzan23}.
% ; schedule-independent safety essentially reduces the problem to a single equivalence class.
This approach has its origins in the work of \citet{lipton:movers}, and typically uses some form of analysis to determine when statements in a program commute in order to restructure programs into an equivalent form that reduces the set of possible nondeterministic outcomes~\citep{DBLP:conf/popl/ElmasQT09, DBLP:conf/fmcad/KraglQ21, DBLP:journals/pacmpl/GleissenthallKB19, DBLP:conf/pldi/FarzanKP22}.
For programs satisfying schedule-independent safety, there is effectively only one equivalence class, allowing a user of \angelic to dynamically select one ordering to verify.

\section{Conclusion and Future Work}
\label{sec:conclusion}
Schedule-independent safety captures the essence of why internal determinism simplifies reasoning about parallel programs.
In this paper, we have shown how \musketeer provides an expressive platform for proving that language-based techniques ensure schedule-independent safety, and how \angelic can take advantage of schedule-independent safety.
One limitation of schedule-independent safety is that it is restricted to safety properties.
In future work, it would be interesting to extend \musketeer for proving that liveness properties, such as termination, are also schedule-independent.

\begin{acks}
This work was supported by the \grantsponsor{NSF}{National Science Foundation}{} through grant no.~\grantnum{NSF}{2318722} and \grantnum{NSF}{2319168}. The authors thank the anonymous reviewers of the paper and associated artifact for their feedback.
\end{acks}

\section*{Data Availability Statement}
The \musketeer and \angelic logics,
their soundness proofs, and all our case
studies are mechanized in the Rocq prover using the Iris framework.
This mechanization is recorded in an artifact available on Zenodo~\citep{mechanization}.

\ifappendix
\appendix
\section{A Counter-Example to the Existantial Elimination Rule in Musketeer}
\label[appendix]{appendix:counterexample}
\newcommand{\counterex}{\textsf{unsafe}}
\begin{figure}\small\centering
\begin{align*}
\counterex \;\;\eqdef\;\;&\eletprefix{r}{\eref{\vtrue}}\\
&\epar{(\eset{r}{\vtrue})}{(\eset{r}{\vfalse})};\\
&\eassert{(\eget{r})}
\end{align*}
\captionlabel{An Unsafe Code}{fig:counterexample}
\end{figure}

In this section, we explain in more detail why adding
the standard \SL rule for eliminating an existential to \musketeer
would be unsound.

Consider the $\counterex$ program presented in \Cref{fig:counterexample}.
This program allocates a reference $r$
initialized to $\vfalse$,
then executes in parallel two writes,
the left one setting $r$ to $\vtrue$, and the right one to $\vfalse$.
After the parallel phase, the program asserts that the content of $r$
is $\vtrue$.
This program does not satisfy schedule-independent safety.
Indeed, if the left write is scheduled before the right one,
$r$ will contain $\vfalse$ and the $\textsf{assert}$ will fail.
Hence, \musketeer must reject the $\counterex$ program.

Yet, let us attempt a \musketeer proof, and let us show that
this proof would succeed if the user is able to eliminate an existential without restriction.
Let us attempt to verify the following triple
\[\atriple{\iTrue}{\counterex}{\_}{\iTrue}\]
After the allocation of $r$, we have to verify
\[\atriple{r \mapsto \vfalse}{\epar{(\eset{r}{\vtrue})}{(\eset{r}{\vfalse})};\,\eassert{(\eget{r})}}{\_}{\iTrue}\]
Because the program next performs a concurrent write on the same
location $r$, we have to use an \emph{invariant} in order to share the
ownership of $r$ between the two tasks. Let us pick the invariant $(\exists x.\, r\mapsto x)$.
We now have to verify
\[\atriple{\boxedassert{\exists x.\, r\mapsto x}}{\epar{(\eset{r}{\vtrue})}{(\eset{r}{\vfalse})};\,\eassert{(\eget{r})}}{\_}{\iTrue}\]
Using \RULE{M-Store} and standard rules for invariants, we can easily establish the intermediate triple
$\forall n.\,\atriple{\boxedassert{\exists x.\, r\mapsto x}}{\eset{r}{n}}{\_}{\iTrue}$.
Using this intermediate triple, the fact that we can duplicate invariants and rules \RULE{M-Par} and \RULE{M-Bind},
we are left with proving
\[\atriple{\boxedassert{\exists x.\, r\mapsto x}}{\eassert{(\eget{r})}}{\_}{\iTrue}\]
We apply \RULE{M-Bind} to focus on the sub-expression $(\eget{r})$
with the weakest possible intermediate postcondition $Q'$, that is we instantiate $Q'$ with $(\lambda\_\,\_.\iTrue)$.
The two preconditions of \RULE{M-Bind} are
\[\atriple{\boxedassert{\exists x.\, r\mapsto x}}{\eget{r}}{\_}{\iTrue} \qquad\text{and}\qquad \forall\val.\,\atriple{\iTrue}{\eassert{\val}}{\_}{\iTrue}\]
The right triple immediately follows from \RULE{M-Assert}.
Using the opening rule of invariants on the left triple, we have to verify
\[\atriple{\exists x.\, r\mapsto x}{\eget{r}}{\_}{\exists x.\, r\mapsto x}\]
In order to conclude, one has to eliminate the existential, before using \RULE{M-Load}.
Thankfully, \musketeer prevents the user from eliminating the existential,
and the proof ultimately fails.
What happened here? The existential quantification on $x$ hid
the fact that there are two possible \emph{scheduling-dependent} witnesses ($\vtrue$ and $\vfalse$).

\section{Definition of the Similarity between Typing and Shape Environments}
\label[appendix]{appendix:similar}
\begin{figure}\small\centering
% Fixpoint similar τ1 τ2 : Prop :=
%   match τ1,τ2 with
%   | TUnit, TUnit | TInt, TInt | TBool, TBool | THashSet _, THashSet _ => True
%   | TProd x1 x2, TProd y1 y2 => similar x1 y1 /\ similar x2 y2
%   | TArrow x1 x2, TArrow y1 y2 => x1=y1 /\ x2=y2
%   | TRef _,TRef _ | TEmpty, TEmpty | TIntArray _, TIntArray _ => True
%   | TPRead q1, TPRead q2 | TPWrite q1, TPWrite q2
%   | TPRead q1, TPWrite q2 | TPWrite q1, TPRead q2 => True
%   | _,_ => False
%   end.
\begin{mathpar}
\inferrule[S-Empty]{}{\similartyp{\tempty}{\tempty}}

\inferrule[S-Unit]{}{\similartyp{\tunit}{\tunit}}

\inferrule[S-Bool]{}{\similartyp{\tbool}{\tbool}}

\inferrule[S-Int]{}{\similartyp{\tint}{\tint}}

\inferrule[S-HashSet]{}{\similartyp{\tintset{\qp_1}}{\tintset{\qp_2}}}

\inferrule[S-Array]{}{\similartyp{\tintarray{\qp_1}}{\tintarray{\qp_2}}}

\inferrule[S-Ref]{}{\similartyp{\tref{\tau_1}}{\tref{\tau_2}}}

\inferrule[S-Arrow]{}{\similartyp{\tarrow{\tau_1}{\tau_2}}{\tarrow{\tau_1}{\tau_2}}}

\inferrule[S-Prod]
{\similartyp{\tau_1}{\tau_1'} \\
\similartyp{\tau_2}{\tau_2'}
}
{\similartyp{\tprod{\tau_1}{\tau_2}}{\tprod{\tau_1'}{\tau_2'}}}

\inferrule[S-PRead]
{\tau = \tpread{\qp_2} \;\;\lor\;\;\tau = \tpwrite{\qp_2}}
{\similartyp{\tpread{\qp_1}}{\tau}}

\inferrule[S-PWrite]
{\tau = \tpread{\qp_2} \;\;\lor\;\;\tau = \tpwrite{\qp_2}}
{\similartyp{\tpwrite{\qp_1}}{\tau}}
\end{mathpar}
\captionlabel{Similar Predicate on Types}{fig:similartyp}
\end{figure}

\begin{figure}
% Fixpoint similar_shape s1 s2 : Prop :=
%   match s1,s2 with
%   | SProd s11 s12, SProd s21 s22 => similar_shape s11 s21 /\ similar_shape s12 s22
%   | SNone, SNone | SRef _ _, SRef _ _ | SArray _, SArray _   | SHashSet _, SHashSet _ => True
%   | SRPrio _, SRPrio _ | SWPrio _, SWPrio _ | SWPrio _, SRPrio _ | SRPrio _, SWPrio _ => True
%   | SArrow g1, SArrow g2 => g1=g2
%   | SInvalid, SInvalid => True
%   | _,_ => False end.
\begin{mathpar}
\inferrule[S-SNone]{}{\similarshape{\snone}{\snone}}

\inferrule[S-SHashSet]{}{\similarshape{\sintset{X_1}}{\sintset{X_2}}}

\inferrule[S-SArray]{}{\similarshape{\sintarray{\vals_1}}{\sintarray{\vals_2}}}

\inferrule[S-SRef]{}{\similarshape{\sref{\val_1}}{\tref{\val_2}}}

\inferrule[S-SArrow]{}{\similarshape{\sarrow{\gamma}}{\sarrow\gamma}}

\inferrule[S-SProd]
{\similarshape{\shape_1}{\shape_1'} \\
\similarshape{\shape_2}{\shape_2'}
}
{\similarshape{\sprod{\shape_1}{\shape_2}}{\sprod{\shape_1'}{\shape_2'}}}

\inferrule[S-SPRead]
{\shape = \spread{\ofs_2} \;\;\lor\;\;\shape = \spwrite{\ofs_2}}
{\similarshape{\spread{\ofs_1}}{\shape}}

\inferrule[S-SWrite]
{\shape = \spread{\ofs_2} \;\;\lor\;\;\shape = \spwrite{\ofs_2}}
{\similarshape{\spwrite{\ofs_1}}{\shape}}
\end{mathpar}
\captionlabel{Similar Predicate on Shapes}{fig:similarshape}
\end{figure}

\Cref{fig:similartyp}
shows the definition of $\similartyp{\tau_1}{\tau_2}$,
asserting that the two \minidet types $\tau_1$ and $\tau_2$ are similar~(\cref{sec:toy}).
This property asserts that both types have the same structure
except that functions must be equal and
that priority reads and priority writes are identified.
\Cref{fig:similarshape}
shows the definition of $\similarshape{\shape_1}{\shape_2}$,
asserting that the two shapes $\shape_1$ and $\shape_2$ are similar.
This property asserts that both shapes have the same structure,
except that function shapes must be equal
and that priority reads and priority writes are identified.

We extend these two predicates to maps $m_1 \approx m_2$
as the trivial predicate for the keys not in the intersection
of $\dom{m_1}$ and $\dom{m_2}$
and the similar predicate when a key appears in both maps.

\section{Additional Explanations on the Concurrent Hash Set Example}

\subsection{A Typing Rule for parfor}
\label[appendix]{appendix:hash:parfor}
\begin{figure}\centering\small
\begin{mathpar}
\inferrule[T-ParFor]
{\Gamma(x)= \Gamma(y) = \tint \\
\fractional{\Gamma}{\Gamma_f} \\
\forall\qp.\;\judg{\minsert{i}{\tint}(\Gamma_f\,\qp)}{\expr}{\tunit}{\Gamma_f\,\qp}
}
{\judg{\Gamma}{\parfor{x}{y}{(\lambda i.\;\expr)}}{\tunit}{\Gamma}}
\end{mathpar}
\captionlabel{A \minidet Type for parfor}{fig:parfortype}
\end{figure}

In order to give a type in \minidet to $\dedupname$~(\cref{sec:dedup}),
we first give
$\parforname$ a type, which we prove sound by dropping
to the semantic model.
\RULE{T-ParFor}, which appear in \Cref{fig:parfortype},
requires the two indices to be variables bound to integers,
for simplicity.
% \am{this is unsatisfactory, but functions with multiple arguments are really tedious.
% I should have backed multi-arguments function in the language... }
It then requires the environment $\Gamma$ to be \emph{fractional},
that is, to contain only fractional assertion. This is witnessed
by the precondition $\fractional{\Gamma}{\Gamma_f}$
which is defined as $(\forall n.\, n \neq 0 \implies \Gamma = \cdot^{n}(\Gamma_f\,n))$,
that is, for every positive integer $n$, $\Gamma_f\,n$ represents a n-th share of $\Gamma$.
Finally, \RULE{T-ParFor} requires to type the last argument of
$\parforname$, which must be a function of the form $\lambda\ofs.\,\expr$.
The precondition requires that $\expr$ is typeable while borrowing a share
$\Gamma_f\,n$ of the environment.

\subsection{Angelic Reasoning Rules for our Concurrent Hash Set}
\label[appendix]{appendix:hash:angelic}
\begin{figure}\centering\small\morespacingaroundstar
\begin{mathpar}
\inferrule[A-HAlloc]
{\always(\forall x.\,\run{(h\,x)}{\lambda\val.\pure{\val=hash\,x}})}
{\run{(\hinit{h}{i})}{\lambda\val.\;\ahashset{\ofs}{\val}{\emptyset}}}

\inferrule[A-HAdd]
{\pure{size\,X < i} \\ \ahashset{i}{v}{X}}
{\runex{(\hadd{v}{x})}{\wal}{\pure{\wal=\vunit} \star \ahashset{\ofs}{\val}{(\singleton{x} \cup X)}}}

\inferrule[A-HElems]
{\ahashset{i}{v}{X}}
{\runex{(\helems{v})}{\val'}{\exists \loc\,\wals.\,\pure{\val'=\loc} \star \loc \mapsto \wals \star \pure{\deduped{X}{\wals}}}}
\end{mathpar}
\captionlabel{Angelic Specifications for a Concurrent Hash Set}{fig:angelichashset}
\end{figure}

\Cref{fig:angelichashset} presents the \angelic reasoning rules
for our councurrent hash set~(\cref{sec:hash}).
These specifications involve a representation predicate
$\ahashset{i}{v}{X}$, where $i$ is the capacity (that is, the maximum number that can be contained in the set),
$\val$ is the hash set and $X$ the logical set with the inserted element.
Note that this predicate is \emph{not} fractional, as there is no need to ever split it.

\RULE{A-HAlloc} verifies $\hinit{h}{i}$.
The precondition requires that the hash function $h$ implements
a hash function in the meta-logic.

\RULE{A-HAdd} verifies $\hadd{\val}{x}$.
The precondition requires that $\val$ is a set with content $X$ and capacity $i$.
The user must ensure that the size of the set $X$ is less than the capacity,
in order to guarantee termination.
The postcondition returns the set with an updated model.

\RULE{A-HElems} verifies $\helems{\val}$.
The precondition requires $C_3\,i$ that $\val$ is a set with content $X$.
The postcondition returns a fresh array $\loc$ pointing to $\wals$
such that $\deduped{X}{\wals}$ holds.

\fi

% english.bib is Francois' bibliography, that I really like.
% Please add things that are not in english.bib into local.bib
\bibliography{english,local}

\end{document}